\documentclass[aps,prd,twocolumn,longbibliography]{revtex4-1}

\usepackage[dvipdfmx]{graphicx}
\usepackage{amssymb,amsmath,color}
\usepackage{bm}
\allowdisplaybreaks[1]

\def\ket#1{|#1\rangle}

\def\bracketii#1#2#3{\langle #1|#2|#3 \rangle}

\def\bbra#1{\left[ #1 \right]}
\def\mbra#1{\left\{ #1 \right\}}
\def\sbra#1{\left( #1 \right)}

\begin{document}

\title{Magnetic-field-induced tunability of spin Hamiltonians: \\Resonances and Efimov states in Yb$_2$Ti$_2$O$_7$}

\author{Yasuyuki~Kato$^{1}$, Shang-Shun~Zhang$^{2}$, Yusuke~Nishida$^{3}$,  and C. D. Batista$^{2,4}$}

\affiliation{$^1$ Department of Applied Physics, the University of Tokyo, Tokyo 113-8656, Japan}
\affiliation{$^2$ Department of Physics and Astronomy, University of Tennessee, Knoxville, Tennessee 37996-1200, USA}
\affiliation{$^3$ Department of Physics, Tokyo Institute of Technology, Ookayama, Meguro, Tokyo 152-8551, Japan,}
\affiliation{$^4$ Neutron Scattering Division and Shull-Wollan Center, Oak Ridge National Laboratory, Oak Ridge, Tennessee 37831, USA}

\begin{abstract}
Universality is a powerful concept that arises from the divergence of a characteristic length scale. For condensed matter systems, this length scale is typically the correlation length, which diverges at critical points separating two different phases.  Few-particle systems exhibit a simpler form of universality when the $s$-wave scattering length diverges. A prominent example of universal phenomena is the emergence of an infinite tower of three-body bound states obeying discrete scale invariance, known as the Efimov effect, which has been subject to extensive research in chemical, atomic, nuclear and particle physics. In principle, these universal phenomena can also emerge in the excitation spectrum of condensed matter systems, such as quantum magnets~[Y. Nishida, Y. Kato, and C. Batista, Nat. Phys. {\bf 9}, 93 (2013)]. However, the limited tunability of the effective inter-particle interaction relative to the kinetic energy has precluded so far their observation.  Here we demonstrate that a high degree of magnetic-field-induced  tunability can also be achieved in quantum magnets with strong spin-orbit coupling: a two-magnon resonance condition can be achieved in Yb$_2$Ti$_2$O$_7$ with a field of $\sim$ 13~T along the [110] direction, which leads to the formation of Efimov states in the three-magnon spectrum of this material. Raman scattering experiments can reveal the field-induced two-magnon resonance, as well as the Efimov three-magnon bound states that emerge near the resonance condition.
\end{abstract}

%75.10.Jm	Quantized spin models, including quantum spin frustration
%75.30.Ds	Spin waves
%03.65.Ge	Solutions of wave equations: bound states
%03.65.Nk	Scattering theory

\maketitle

\section{Introduction}

The simplest example of universality arises in the vicinity of scattering resonances of few-body systems,  where the low-energy physics is characterized solely by the $s$-wave scattering length $a$. One of the most prominent observations of universal phenomena at $a\to\infty$ is the emergence of an infinite tower of three-body bound states obeying discrete scale invariance, known as the Efimov effect~\cite{Efimov:1970}:
\begin{align}\label{eq:scaling}
 \frac{E_{n+1}}{E_n} \to \lambda^{-2} \qquad (n\to\infty)
\end{align}
with the universal scale factor $\lambda=22.6944$. For the last five decades, this effect has been subject to extensive research in chemical, atomic, nuclear and particle physics~\cite{Nielsen:2001,Braaten:2006,Ferlaino:2010,Hammer:2010,Naidon:2017,Greene:2017,D_Incao:2018}.
However, Efimov states have been observed in very limited systems due to the requirement of proximity to a scattering resonance. Thus, tunability of inter-particle interactions are highly desired for their realization, which has only been achieved for atomic gases. As we will demonstrate here, this tunability can also be achieved in quantum magnets with strong spin-orbit coupling, which opens the possibility of studying and observing Efimov states in condensed matter systems.

In general, the main obstacle for observing the universality in the vicinity of scattering resonances in condensed matter systems is their limited tunability in comparison to ultracold atoms, whose Feshbach resonances provide a way to vary $a$ by applying a uniform magnetic field~\cite{Cheng:2010}. Among multiple uses, this tool has served to study the crossover between Bose-Einstein condensates (BECs) of fermionic molecules and  the BCS regime of weakly interacting fermion-pairs in Fermi clouds~\cite{Regal:2004,Zwierlein:2004,Kinast:2004,Bourdel:2004,Chin:2004,Partridge:2005,Zwierlein:2005}. For the BECs, Feshbach resonances have been used to study a variety of systems from the non-interacting ideal Bose gases to the unitary regime of interactions~\cite{Navon:2011,Rem2013,Fletcher:2013,Makotyn:2014,Eismann:2016,Fletcher:2017,Klauss:2017,Eigen:2017,Fletcher:2018,Eigen:2018}. Is it then possible to find a counterpart  of the Feshbach resonances in solid state physics?

In this paper, we provide an affirmative answer to this question by demonstrating that a uniform magnetic field can also be used to tune the $s$-wave scattering length for the collision between magnons of quantum magnets with strong spin-orbit coupling. 
This goal is achieved by tuning the effective magnon tunneling with the external magnetic field, while keeping the attractive magnon-magnon interaction practically unchanged. This tunability makes it possible to drive the system into its universal regime by approaching the resonance condition.
Because magnons obey bosonic statistics, this is enough to realize  Efimov states in the three-magnon spectrum of quantum magnets~\cite{Nishida:2013}, such as Yb$_2$Ti$_2$O$_7$, as well as other consequences of the universality.

\begin{figure}[t]
  \centering
  \includegraphics[trim=0 0 0 0, clip,width=\columnwidth]{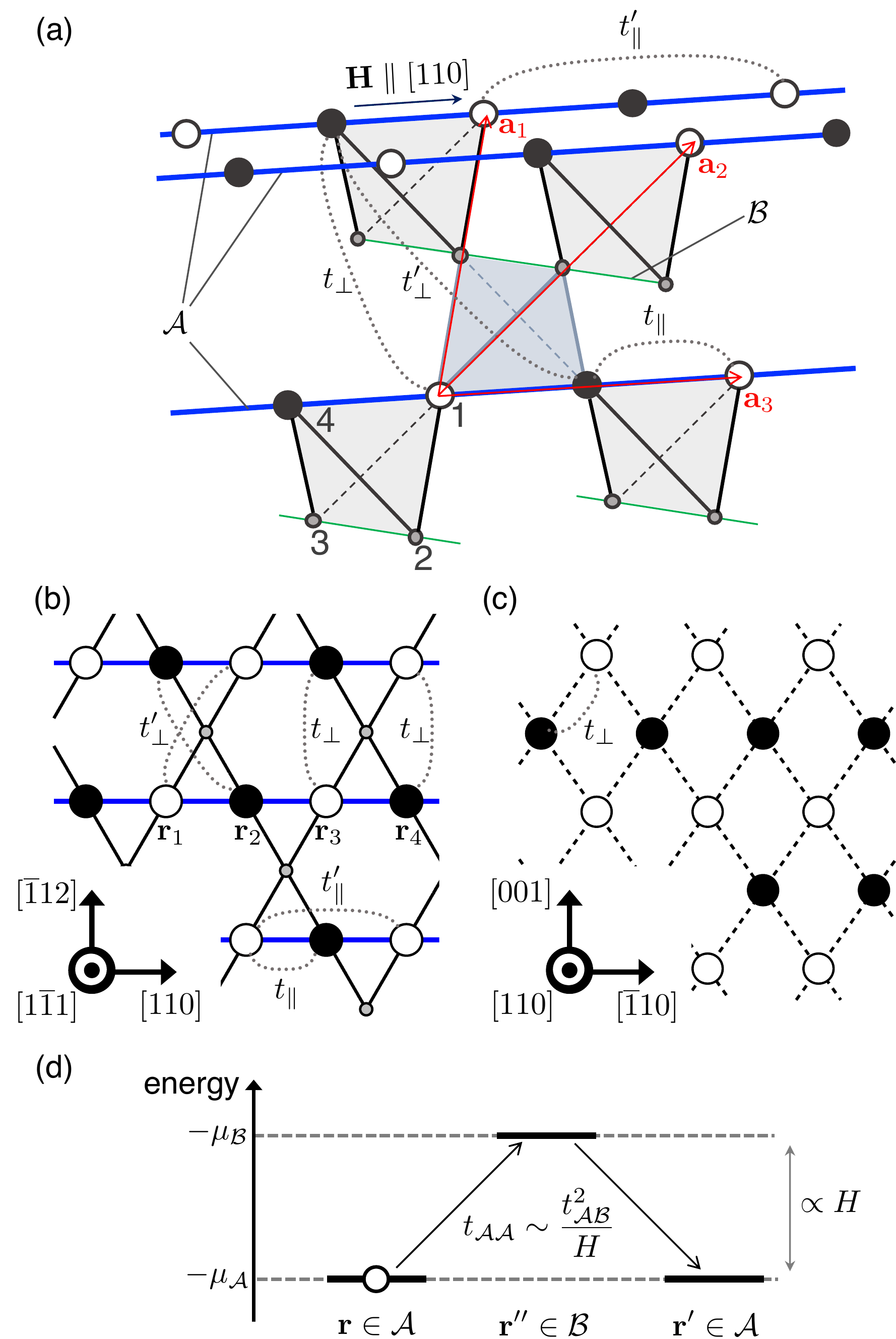}
  \caption{
  Pyrochlore lattice formed by Yb cations.
  (a) Circles represent Yb cations.  
  Primitive translational vectors, ${\bf a}_{1\text{--}3}$, sublattice index, $\alpha=1\text{--}4$, and the external field direction 
  ${\bf H}\parallel [110]$. 
  Thick blue lines and thin green lines indicate the low-energy $\mathcal{A}$ chains of $\alpha=1$ and $4$ and the high-energy $\mathcal{B}$ chains of $\alpha=2$ and $3$, respectively.
  (b) $(1\bar{1}1)$-plane of the lattice. 
  (c) $(110)$-plane  of the lattice. 
  Hopping paths of $t_\parallel$, $t_{\parallel}'$, $t_\perp$, and $t_\perp'$ in $\mathcal{H}_{\rm eff}$ are indicated. 
  The site indices ${\bf r}_{1\text{--}4}$ represent  an example of $\langle {\bf r}_1{\bf r}_2{\bf r}_3 \rangle$ or $\langle {\bf r}_1{\bf r}_2{\bf r}_3{\bf r}_4 \rangle$ in $\mathcal{H}_{\rm eff}$.
  (d) Schematic picture of effective hopping $t_{\mathcal{AA}}$ induced by $t_{\mathcal{AB}}$. 
  $\mu_{\mathcal{A}}$ and $\mu_{\mathcal{B}}$ represent the chemical potential of the low-energy and high-energy sublattices, respectively.
  }
  \label{fig:lattice}
\end{figure}

Similar to the case of atomic gases, an external magnetic field works as an effective chemical potential for the magnons of the fully polarized magnetic ground state that is induced above the saturation field~\cite{Zapf:2014}. 
A key observation here is that the chemical potential  can be made inhomogeneous in magnets with strong spin-orbit coupling and more than one magnetic ion per unit cell. For instance,  Yb$_2$Ti$_2$O$_7$ comprises a  pyrochlore lattice (Fig.~\ref{fig:lattice}) of Yb$^{3+}$ cations
that can be divided into four symmetry related sublattices, 1, 2, 3 and 4 corresponding to the four corners of each tetrahedron with local high-symmetry axes $[111]$,  $[1 {\bar 1} {\bar 1}]$, $[{\bar 1} 1 {\bar 1}]$ and $[{\bar 1} {\bar 1} 1]$ respectively.
Because the effective $g$-tensor of each magnetic ion is strongly anisotropic, it has a strong sublattice dependence in a global reference frame. In other words, the chemical potential induced by an external field  ${\bf H}=(H^x,H^y,H^z)$ is sublattice dependent.
For $ {\bf H} \parallel [110] $, the four sublattices are divided into two pairs: the low-energy $\mathcal{A}$ sublattices 1 and 4
with  chemical potential $\mu_\mathcal{A}$ and the high-energy $\mathcal{B}$ sublattices 2 and 3  
with chemical potential $\mu_\mathcal{B}$. As it is indicated in Fig.~\ref{fig:lattice}(a), the magnetic ions in the $\mathcal{A}$ ($\mathcal{B}$) sublattices form chains running along the $[110]$ ($[1{\bar 1}0]$) direction.
Because $\mu_\mathcal{A} \gg \mu_\mathcal{B}$ for large enough magnetic field values, the 
$\mathcal{A}$ ($\mathcal{B}$) chains become low-energy (high-energy) chains for $H\equiv |{\bf H}| \to \infty$.
Given that the $\mathcal{A}$ and $\mathcal{B}$ sublattices form a bipartite graph ({\it bare} magnon tunneling  $t_{\mathcal{AB}}$ only exists between $\mathcal{A}$ and $\mathcal{B}$ sublattices), the {\it effective} magnon tunneling  $t_{\mathcal{AA}}$ between different  low-energy chains
can be continuously suppressed by increasing the energy difference $|\mu_\mathcal{A} - \mu_\mathcal{B}|$ [Fig.~\ref{fig:lattice}(d)].
Since $\mu_\mathcal{A} - \mu_\mathcal{B}$  is roughly proportional to  $H$ and $t_{\mathcal{AA}} \sim t_{\mathcal{AB}}^2/H$, the field can be used  to vary the effective magnon tunneling between different low-energy chains. In particular, the original  three-dimensional system becomes quasi-one-dimensional in the large field limit.
While the above-described set up is the one that will be used in this paper,  we note that it is also possible to make the system quasi-two-dimensional by applying the field along the $[111]$ direction. In this case, the $\mathcal{A}$ and $\mathcal{B}$ subsystems correspond to alternating triangular and Kagome layers, respectively.

%%%%%%%%%%%%%%%%%%%%%%%

The structure of this paper is as follows. 
In Sec.~\ref{sec:model_method}, we introduce the low-energy effective hard-core boson model derived from the effective spin model of Yb$_2$Ti$_2$O$_7$. 
In addition, we describe the numerical calculation methods to solve the two- and three-magnon problems.
In Sec.~\ref{sec:results}, we show the results of analytical and numerical calculations of the one-, two-, and three-magnon problems. 
We demonstrate that the magnon scattering length can be tuned with an external magnetic field producing a two-magnon resonance condition for a field strength of $\sim$ 13~T along the [110] direction. We also demonstrate that  Efimov states emerge in the three-magnon sector near the two-magnon resonance condition. Finally, in Sec.~\ref{sec:summary_discussions}, we summarize the main results and discuss their experimental realization.
All technical details are presented in Appendices.

\section{Model and Method}\label{sec:model_method}

\subsection{Effective hard-core boson model\label{sec:model}}

Yb$_2$Ti$_2$O$_7$ comprises a pyrochlore lattice of magnetic Yb$^{3+}$ ions [Fig.~\ref{fig:lattice}(a)], whose low-energy degrees of freedom (doublets) are described by an effective spin 1/2 Hamiltonian~\cite{Onoda:2011,Lee:2012}: 
\begin{align}
\mathcal{H}_{\rm spin} =& 
\sum_{\langle {\bf r}{\bf r}' \rangle} \large[
J_{zz} \mathsf{S}^z_{\bf r} \mathsf{S}^z_{{\bf r}'}  
-J_{\pm} \left(
\mathsf{S}^+_{\bf r} \mathsf{S}^-_{{\bf r}'}  
+ {\rm h.c.}
\right) \nonumber \\
&+J_{\pm\pm}
\left(
\gamma_{\alpha_{\bf r} \alpha_{{\bf r}'}  } \mathsf{S}^+_{\bf r} \mathsf{S}^+_{{\bf r}'}  
+ {\rm h.c.}
\right) \nonumber \\
&+J_{z \pm}
\left\{
\mathsf{S}^z_{{\bf r}}  
\left(
\zeta_{\alpha_{\bf r} \alpha_{{\bf r}'}  } \mathsf{S}^+_{{\bf r}'}  
+ {\rm h.c.}
\right)+
({\bf r} \leftrightarrow {\bf r}')
\right\}
\large] \nonumber \\
&-\mu_{\rm B} \mu_0 \sum_{\eta, \nu} H^{\eta} \sum_{\bf r} g^{\eta \nu}_{\alpha_{{\bf r}}  } \mathsf{S}^\nu_{{\bf r}}.
\label{eq:Hspin}
\end{align}
The spin 1/2 operators, $\mathsf{S}_{\bf r}^\nu$ ($\nu = x$, $y$, or $z$, and $\mathsf{S}^{\pm}_{\bf r} = \mathsf{S}^x_{\bf r} \pm {\rm i} \mathsf{S}^y_{\bf r}$), are expressed in a sublattice dependent reference frame, whose local $z$-axis is parallel to the local [111] direction. 
The index $\alpha_{\bf r} =$ 1--4 indicates the sublattice of the site ${\bf r}$, and
$\langle {\bf r}{\bf r}' \rangle$ indicates that the sum $\sum_{\langle {\bf r}{\bf r}' \rangle}$ runs over the nearest-neighbor sites of the pyrochlore lattice.
The sums of $\eta$ and $\nu$ run over $x$, $y$, and $z$.
 $\mu_{\rm B}$ is  the Bohr magneton, %%$\mu_{\rm B} = 5.7883818012(26) \times 10^{-2} ~({\rm meV/T})$
and $\mu_0$ is the permeability constant.
 $\gamma_{\alpha_{\bf r} \alpha_{{\bf r}'}}$ and  $\zeta_{\alpha_{\bf r} \alpha_{{\bf r}'}}$ are phase factors,
 and $g_{\alpha_{\bf r}}$ is the $g$-tensor for sites in the sublattice $\alpha_{\bf r}$ (Appendix~\ref{sec:spin_hamiltonian}). 
 The model parameters
($J_{zz}$,
$J_{\pm\pm}$,
$J_{\pm}$,
$J_{z\pm}$,
$g_{\parallel}$,
and
$g_{\perp}$)
are set to the values estimated from a recent inelastic neutron scattering experiment of Yb$_2$Ti$_2$O$_7$~\cite{Thompson:2017}. The effective spin 1/2 moments can be mapped into hard core bosons with creation and annihilation
operators $a^\dag_{\bf r}$ and $a^{\;}_{\bf r}$, respectively~\cite{Matsubara:1956}. 
By choosing the local quantization axis to be parallel to the magnetic moment,
the exact mapping leads to a hard-core boson model with no linear terms in the creation or annihilation operators (Appendix~\ref{sec:hardcoreboson_hamiltonian}).  

The Zeeman term, which is dominant for relatively high fields ($\mu_0 H \gg |J_{zz}|, |J_{\pm}|, |J_{\pm\pm}|, |J_{z\pm}|$), becomes a sublattice  dependent chemical potential term in the hard-core boson language, 
with $|\mu_\alpha|$ being roughly proportional to $H$. In particular, the difference between $\mu_1=\mu_4$ and $\mu_2 = \mu_3$ increases in proportion to $H$ because of the strongly  anisotropic $g$-tensor that arises from a combination of strong spin-orbit coupling and the crystal field of the Yb cation. The energy scale of all the other 
terms in Hamiltonian is much smaller than  $|\mu_\alpha |$ and $| \mu_1 - \mu_2  |$ for high enough fields. Thus, we will regard the chemical potential (Zeeman) terms as the unperturbed Hamiltonian and we will treat the rest as perturbation.

By applying the second order degenerate perturbation theory  (Appendix~\ref{sec:derivation_Heff}), we obtain an effective low-energy 
Hamiltonian for bosons on the  chains 1 and 4 that conserves the particle number:
\begin{align}
\mathcal{H}_{\rm eff}
&=-\mu\sum_{\bf r} n_{{\bf r}}  \nonumber\\
&+t_\parallel \sum_{\langle {\bf r}{\bf r}' \rangle_\parallel} (a^\dag_{\bf r} a^{\;}_{{\bf r}'}  +{\rm h.c.})
+t'_\parallel \sum_{\langle {\bf r}{\bf r}' \rangle'_\parallel} (a^\dag_{\bf r} a^{\;}_{{\bf r}'}  +{\rm h.c.})\nonumber \\
&+t_\perp \sum_{\langle {\bf r}{\bf r}' \rangle_\perp} (a^\dag_{\bf r} a^{\;}_{{\bf r}'}  +{\rm h.c.})
+t'_\perp \sum_{\langle {\bf r}{\bf r}' \rangle'_\perp} (a^\dag_{\bf r} a^{\;}_{{\bf r}'}  +{\rm h.c.})\nonumber \\
&+u_\parallel \sum_{\langle {\bf r}{\bf r}' \rangle_\parallel} n_{\bf r} n_{{\bf r}'}  
+u'_\parallel \sum_{\langle {\bf r}{\bf r}' \rangle'_\parallel} n_{\bf r} n_{{\bf r}'}  
+u_\perp \sum_{\langle {\bf r}{\bf r}' \rangle_\perp}  n_{\bf r} n_{{\bf r}'}  \nonumber\\
&+u'_\perp \sum_{\langle {\bf r}{\bf r}' \rangle'_\perp}  n_{\bf r} n_{{\bf r}'}  
+v_1\sum_{\langle {\bf r}_1{\bf r}_2{\bf r}_3 \rangle} (a^\dag_{{\bf r}_1} n^{\;}_{{\bf r}_2} a^{\;}_{{\bf r}_3} +{\rm h.c.}) \nonumber\\
&+v_2\sum_{\langle {\bf r}_1{\bf r}_2{\bf r}_3{\bf r}_4 \rangle} (a^\dag_{{\bf r}_2}   a^{\;}_{{\bf r}_3} +{\rm h.c.})(n^{\;}_{{\bf r}_1} +n^{\;}_{{\bf r}_4}) \nonumber\\
&+w\sum_{\langle {\bf r}_1{\bf r}_2{\bf r}_3 \rangle} n_{{\bf r}_1} n_{{\bf r}_2}  n_{{\bf r}_3}
+ \mathsf{U} \sum_{\bf r} n_{\bf r} (n_{\bf r} -1),
\label{eq:boson_mdl}
\end{align}
where $t'_\perp = t_\perp$ and $u'_\perp = - u_\perp$. $\mu$ is the chemical potential including the second order correction, the following four hopping terms represent the kinetic energy, and the rest of the terms are multi-body interactions. The brackets  $\langle {\bf rr}^{\prime} \rangle_{\parallel}$, $\langle {\bf rr}^{\prime} \rangle_{\parallel}'$, $\langle {\bf rr}^{\prime} \rangle_{\perp}$, and $\langle {\bf rr}^{\prime} \rangle_{\perp}'$ indicate  that the sums run over intrachain nearest-neighbors (n.n.), intrachain next n.n., interchain n.n., and interchain next n.n. separated by a site on sublattices 2 or 3, respectively [Figs.~\ref{fig:lattice}(a) and \ref{fig:lattice}(b)]. The brackets $\langle {\bf r}_1{\bf r}_2 {\bf r}_3 \rangle$ and $\langle  {\bf r}_1{\bf r}_2 {\bf r}_3 {\bf r}_4 \rangle$ indicate that the corresponding sums  run over all possible combinations of consecutive  three and four sites, respectively~[Fig.~\ref{fig:lattice}(b)].
$\mathsf{U}(=\infty)$ is the on-site repulsion that enforces the hard-core constraint.

\begin{figure}[t]
  \centering
 \includegraphics[trim=0 0 0 0, clip,width= \columnwidth]{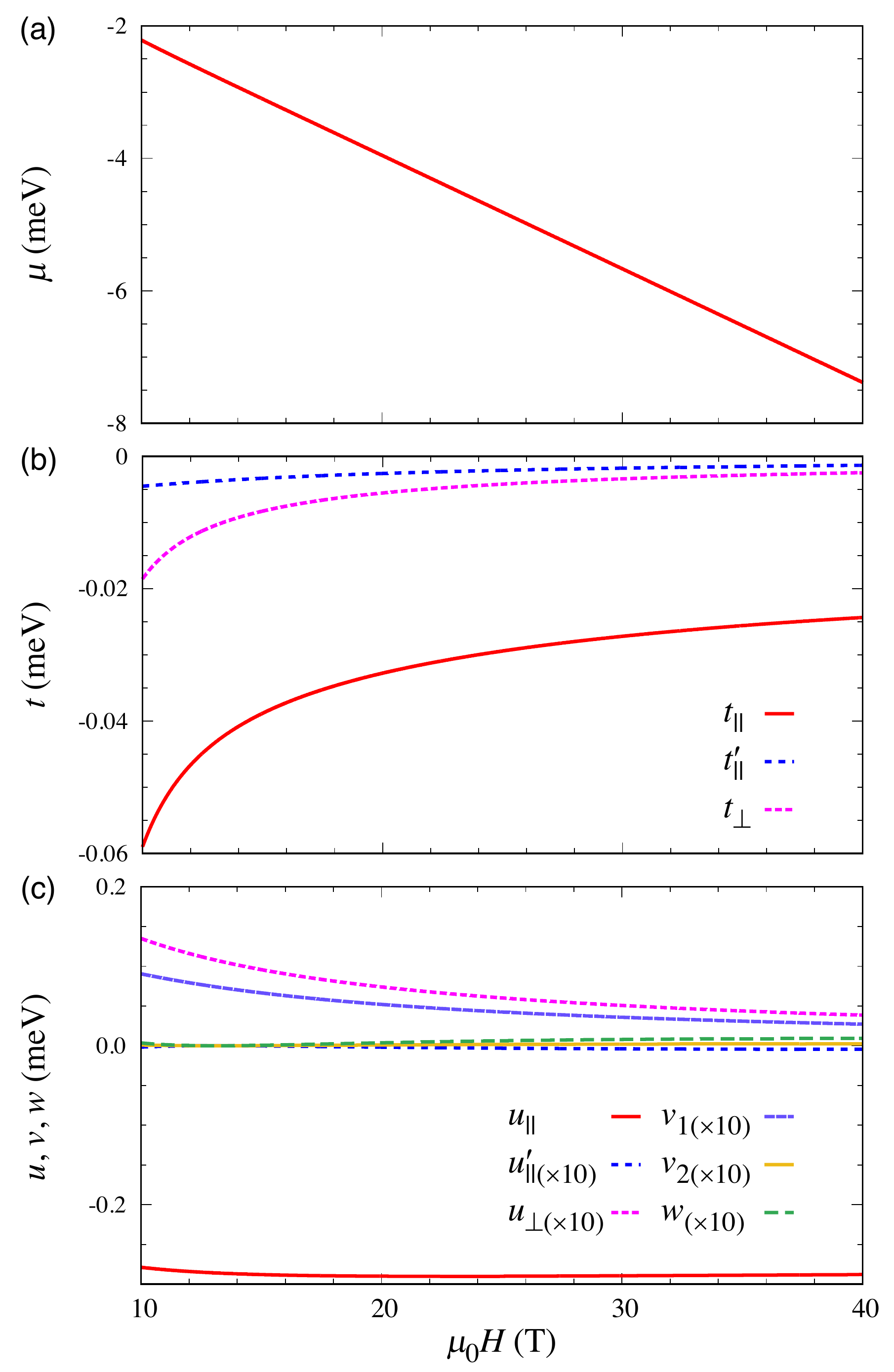}
  \caption{
  Model parameters for the effective boson Hamiltonian
  $\mathcal{H}_{\rm eff}$
 obtained for the spin Hamiltonian 
  $\mathcal{H}_{\rm spin}$  reported in Ref.~\cite{Thompson:2017}.
  	(a) Chemical potential $\mu$.
    	(b) Hopping amplitudes $t_\parallel$, $t_\parallel'$, and $t_\perp$.
	(c) Interactions $u_\parallel$, $u_\parallel'$, $u_\perp$, $v_1$, $v_2$, and $w$.
	For better visibility, all the interactions, except for $u_\parallel$, are multiplied by 10.}
  \label{fig:eff_mdl_param}
\end{figure}

Figure~\ref{fig:eff_mdl_param} shows the field dependence of the parameters of $\mathcal{H}_{\rm eff}$ in the field range of $\mu_0 H=10$--$40$~T
where the energy scale of $\mu$ is larger than those of the other parameters by one order of magnitude.
While all the hopping amplitudes have a strong field dependence, the relative change of the dominant attractive interaction $u_\parallel$ remains very small over the whole field range. In other words, the ratio of the attractive interaction and the kinetic energy is widely tunable by the external field. This behavior resembles the case of ultracold atomic gases trapped in a periodic optical lattice, where the hopping amplitude is controlled by tuning the depth of the periodic potential~\cite{Jaksch:1998}. In our case, however, the strong field dependence of $t_\parallel$ and $t_\perp$ is caused by a different mechanism: the amplitude of the magnon tunneling via the ``high-energy'' sublattices 2 or 3 is inversely proportional to the energy barrier $\mu_1 - \mu_2$.

While the hopping amplitudes are comparable to each other, the interactions $u_{\parallel}'$, $v_2$, and $w$ are much smaller than the other interactions in the field range of $\mu_0 H=10$--$40$~T.
We will then ignore these three interactions hereafter to reduce the computational cost of solving the two-body and three-body problems. Our exact diagonalization results for 
$\mathcal{H}_{\rm eff}$ on a cluster of linear size $L=12$ (i.e., $2\times L^3$ sites)
confirm that  these interactions have indeed a negligible effect.

\subsection{Numerical calculation methods for two- and three-magnon problems}

We numerically analyze the two- and three-magnon  problems using the effective hard-core boson model, in which the number of magnons is conserved. 
The eigenvalues and the eigenstates of the two- and three-magnon sectors of the Hamiltonian \eqref{eq:boson_mdl} 
are obtained from exact diagonalization on finite lattices 
and from a numerical solution of the Lippmann-Schwinger equation.
In the exact diagonalization method, we use the Krylov-Shur algorithm (library SLEPc~\cite{Hernandez:2005}) to compute the lowest energy state in each sector.
The advantage of the exact diagonalization method is its simpler implementation.
The calculations are performed with lattices of linear size $L \leq 72$ and $L \leq 18$ for the two- and three-magnon bound states, respectively.
It is confirmed that these linear sizes are large enough for the accurate estimates of the $s$-wave scattering length and the binding energies of the two-magnon bound state and the lowest three-magnon bound state.
The binding energy of the latter is well converged with respect to the system size, because its linear size is as small as a few lattice spacings.
On the other hand, the same is not true for the first excited three-magnon bound state, because its linear size is larger and comparable to the maximum system size that can be reached with the state of the art exact diagonalization method. However, its binding energy can still be computed by solving the Lippmann-Schwinger equation
with the Gaussian quadrature rule for the numerical integrations in momentum space.

For the solution of the Lippmann-Schwinger equation, we consider essentially the same linear integral equations that were introduced in Ref.~\cite{Nishida:2013} for the case of the simple cubic lattice.
However, the number of equations increases from 2 to  24  because of the multiple sublattice structure of the pyrochlore lattice. Detailed derivations of the Lippmann-Schwinger equations for the two- and three-magnon sectors are given in the Appendix~\ref{sec:solution_of_few_body_problem}.

\section{Results}\label{sec:results}
\subsection{Single-magnon spectrum}

The single-magnon dispersion is obtained by diagonalizing the one-body component (first five terms) of 
$\mathcal{H}_{\rm eff}$
 (Appendix~\ref{sec:single_magnon_problem}). For $t_\parallel$, $t'$, $t_\perp<0$, the lower branch of the spectrum $E_{-}({\bf k})$ has a minimum at ${\bf k}={\bf 0}$
where ${\bf k}=\sbra{ k_1 {\bf G}_1 + k_2 {\bf G}_2 + k_3 {\bf G}_3 }/2\pi $
with the reciprocal lattice vectors, ${\bf G}_{1\text{--}3}$, for the primitive  vectors, ${\bf a}_{1\text{--}3}$, shown in Fig.~\ref{fig:lattice}(a). 
In the long-wavelength limit, $\vert {\bf k} \vert \ll 1 $, we find
\begin{equation}
E_-({\bf k}) \simeq
-\mu+
 2t_\parallel + 2t'_\parallel + 8t_\perp +\frac{\overline{\bf k}^2}{2m_z},
 \label{eq:Disprel}
\end{equation}
where
$\overline{\bf k}= [ \sqrt{{m_z}/{m_x}} ( k_1 - {k_3}/{2} ),
 \sqrt{{m_z}/{m_x}} (k_2 - {k_3}/{2}),
 k_3
 ]$,
$m^{-1}_z  =  - t_\parallel / 2  - 2t'_\parallel - t_\perp$, and
$m^{-1}_x  = - 4t_\perp$.
The effective masses $m_z$ and $m_x$ correspond to the $[110]$ and the ${\bf G}_{1,2}(\perp [110])$ directions, respectively. 
Therefore, the low-energy physics of Yb$_2$Ti$_2$O$_7$ is described by bosons in continuous space with the anisotropic mass tensor.
Two- and three-magnon binding energies discussed below are measured from the bottoms of two- and three-magnon continua at $E = 2E_-({\bf0})$ and $3E_-({\bf0})$, respectively.

%%%%%%%%%%%%%%%%%%
\subsection{Two-magnon resonance}

\begin{figure}[t]
  \centering
 \includegraphics[trim=0 0 0 0, clip,width=\columnwidth]{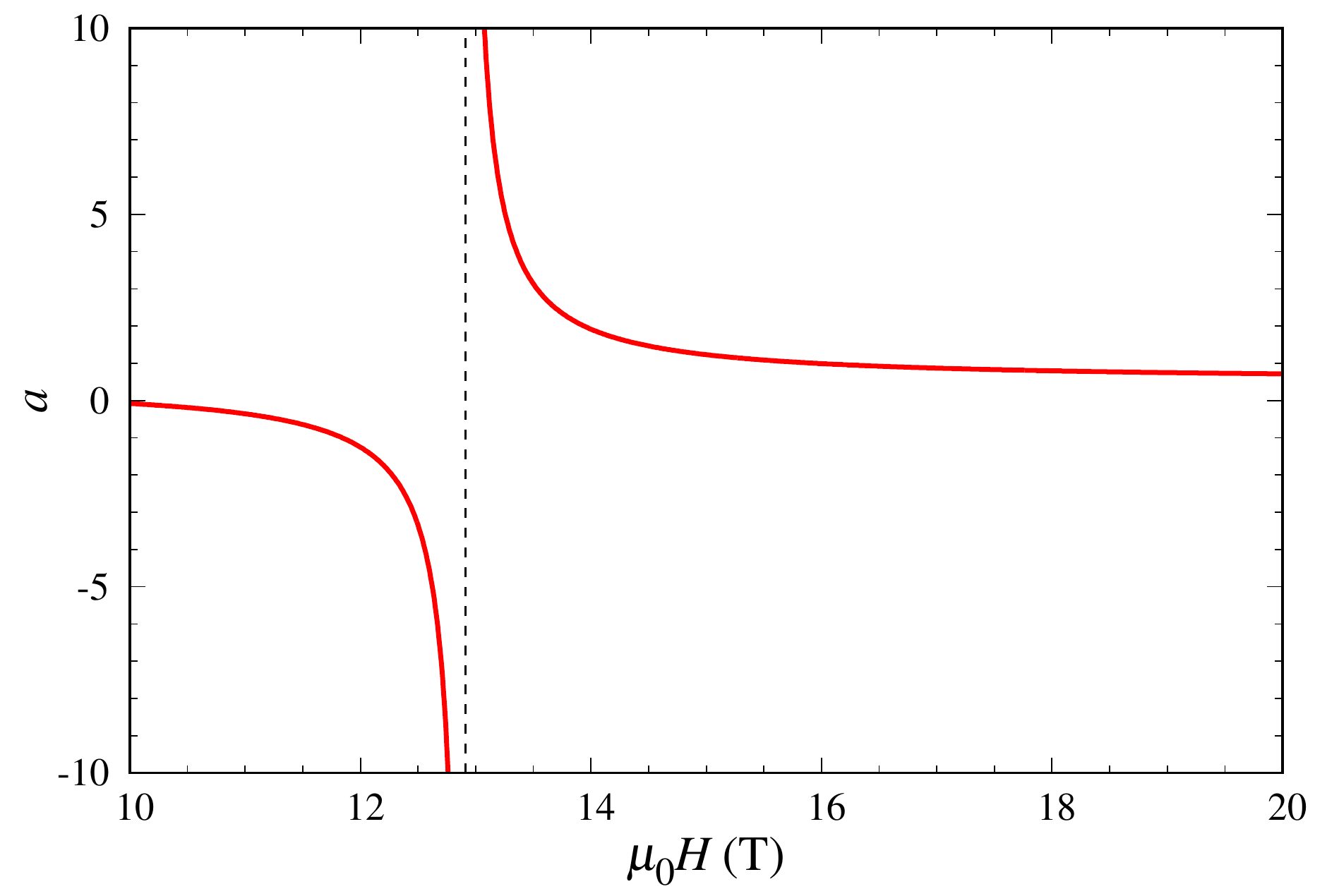}
  \caption{
  $s$-wave scattering length in the unit of lattice spacing. 
  The scattering length $a$ (red line) shows the divergent behavior of 
  $a\propto(H-H_c)^{-1}$ near the two-magnon resonance condition $\mu_0 H_c=12.91$~T (dashed line). 
 }
  \label{fig:result0}
\end{figure}

The low-energy scattering of magnons is parametrized by the $s$-wave scattering length $a$, which can be extracted by solving the two-magnon problem (Appendix~\ref{sec:two_magnon_problem}). Because of the field dependence of the parameters of $\mathcal{H}_{\rm eff}$, $a$ also varies with the magnetic field. 
Figure~\ref{fig:result0} shows the field dependence of $a$, establishing its magnetic-field-induced tunability. 
In particular, we find the divergent behavior of $a$ at $\mu_0 H_c = 12.91$ T, which corresponds to the two-magnon resonance condition and signals the onset of a two-magnon bound state for $H > H_c$. 
The green line in Fig.~\ref{fig:result1} then shows the field dependence of the  two-magnon binding energy that is obtained from the exact diagonalization of 
$\mathcal{H}_{\rm eff}$. As expected from the universality,  the binding energy vanishes as  $1/(m_z a^2)$ upon approaching $H_c$. 
We note that the two-magnon bound state dispersion has a global minimum at  center-of-mass momentum ${\bf K}=0$.

\subsection{Three-magnon Efimov states}

\begin{figure}[t]
  \centering
  \includegraphics[trim=10 0 0 -10, clip,width=\columnwidth]{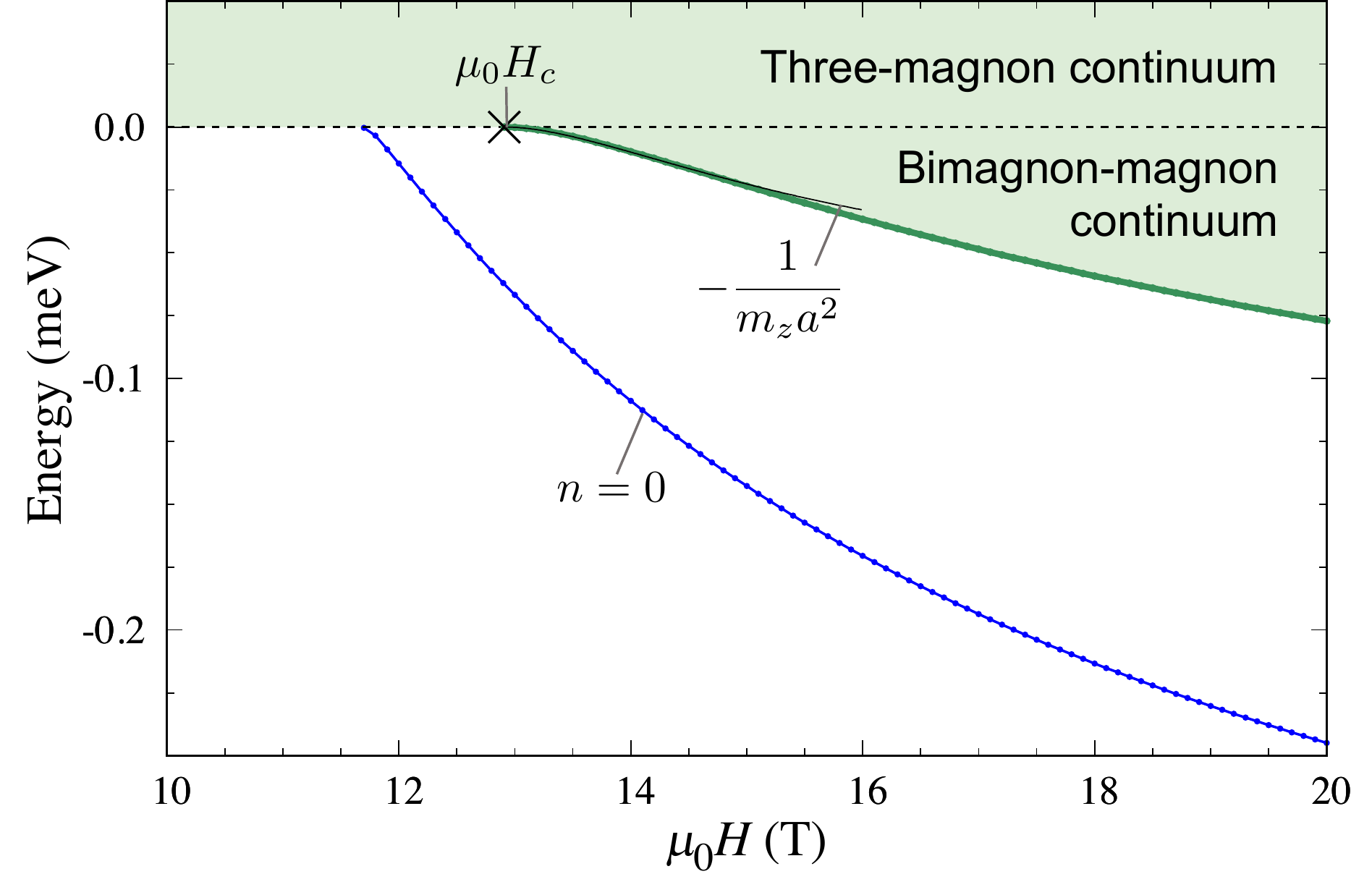}
  \caption{
	Energy spectrum of the three-magnon states measured from $3 E_{-}({\bf 0})$. 
	The cross ($\times$) marks the critical field $\mu_0 H_c$ for the two-magnon resonance.
	The green line represents the lower threshold of the  bimagnon-magnon continuum corresponding to the negative of the two-magnon binding energy.
	The black thin line indicates the universal form of the two-magnon binding energy, $-1/(m_z a^2)$, for $H\to H_c$. 
	The blue circles represent the energy of the lowest ($n=0$) three-magnon bound state
	corresponding to the negative of three-magnon binding energy. 
 }
  \label{fig:result1}
\end{figure}

The exact diagonalization method is also applied to the three-magnon problem to compute the binding energies of the three-magnon bound states.
Since the two-magnon bound state emerges for $H>H_c$, the lower threshold of the continuum is set by an eigenstate consisting of a two-magnon bound state or ``bimagnon'' plus a single magnon. Out of the few three-magnon bound states that appear below this threshold, we can identify two branches of $s$-wave bound states, labeled by $n=0$ and $n=1$, as well as a branch of $p$-wave bound states. 
The $s$-wave bound states are candidates for the Efimov states. The binding energy of the lowest ($n=0$) three-magnon bound state is shown in Fig.~\ref{fig:result1}. The binding energy of the $n=1$ state is not shown there because it is too shallow and hard to distinguish from the two-magnon binding energy (green line).

We then focus on the $s$-wave  three-magnon bound states ($n=0$ and $n=1$) at the resonance condition $\mu_0H = 12.91$~T. 
Our numerical solutions of the Lippmann-Schwinger equation (Appendix~\ref{sec:three_magnon_problem}) produce well converged binding energies for the $n=0$ and $1$ states,
\begin{align}
E_0 =&~ 
0.062\;493~\text{meV},\label{eq:En0}\\
E_1 =&~ 
0.000\;180~\text{meV},\label{eq:En1}
\end{align}
and the square root of their ratio is
\begin{align}
\sqrt{\frac{E_0}{E_1}} = 18.6.
\end{align}
This value deviates from the universal value $\lambda = 22.6944$, for $n\to\infty$,  because  the linear size of the $n=0$ state is comparable to the lattice spacing and lattice effects introduce a significant correction to its universal character. Similar deviations have been reported for the ratio $E_0/E_1$ obtained with a simpler spin Hamiltonian on a cubic lattice~\cite{Nishida:2013}, where the ratios $E_n/E_{n+1}$ for $n=1,2$ are also computed and found to follow the universal value. Due to numerical limitations, we only have access to the $n=0$ and $n=1$ states.
Consequently, to identify the Efimov character of each $n$ state, 
we are led to compare its wave function with the universal wave function of the Efimov state.

It is convenient to express the three-magnon wave function in mixed representation, $\psi_{\bf m}({\bf r};{\bf k} )$, where ${\bf r}$ is the relative coordinate of two magnons and ${\bf k}$ is the relative momentum of the third magnon with respect to the center-of-mass of the other two. ${\bf m} \equiv (m_1, m_2, m_3)$ and $m_{j} = +,-$ corresponding to $\alpha=4,1$, respectively, identifies the sublattice in which the $j$th magnon resides.
The center-of-mass momentum is set to zero because we are interested in the universal behavior that emerges in the long-wavelength limit of the theory.
Similarly to the case of two-body bound states, the three-body bound states are expected to have minimum energy for center-of-mass momentum ${\bf K}=0$ because the single-magnon spectrum has a global minimum at ${\bf k}=0$ [Sec.~IIIA]. 
The three-magnon wave function is obtained by solving the Lippmann-Schwinger equation (Appendix~\ref{sec:three_magnon_problem}).
To compare the resulting wave functions against the universal theory, we express them in terms of the rescaled wave vector $\overline{\bf k}$ introduced in Eq.~\eqref{eq:Disprel}. Here the effective masses of $m_z^{-1}=0.0399$~meV and $m_x^{-1} = 0.0425$~meV for $\mu_0 H =12.91$~T are used. 

\begin{figure}[t]
  \centering
\includegraphics[trim=0 0 0 10,clip,width=\columnwidth]{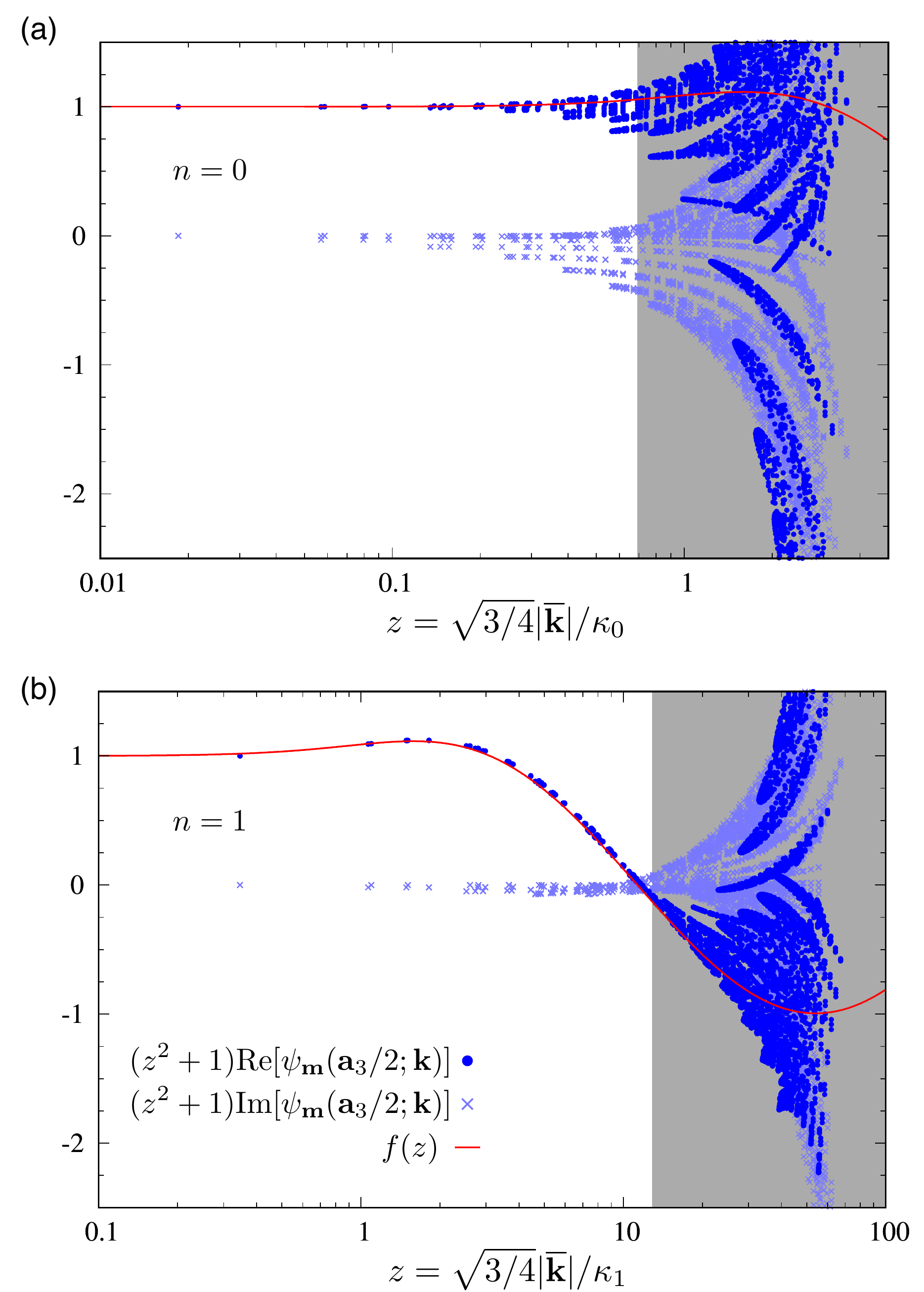}
  \caption{
 Wave functions of the two lowest three-magnon bound states at the critical field $\mu_0H_c = 12.91$ T.
The universal function $f(z)$ and the left hand side of Eq.~\eqref{eq:essential_portion} for (a) the ground state ($n=0$) and for (b) the first excited state ($n=1$) are compared as functions of the rescaled wave number $z = \sqrt{3/4}|\overline{\bf k}|/\kappa_n$ normalized by $\kappa_n$ for each $n$. Here we set the relative coordinate as ${\bf r}={\bf a}_3/2$ so that a pair of particles occupy two nearest-neighbor sites. The gray shaded regions indicate the nonuniversal regime ($|\overline{\bf k}|>1$).
 }
  \label{fig:result2}
\end{figure}

The wave functions of Efimov states obey,
\begin{align}
& \bbra{\frac{3}{4} \sbra{ \frac{|\overline{\bf k}|}{\kappa_n} }^2 + 1} \psi_{\bf m}  ({\bf r};{\bf k})= f\left(\sqrt{\frac34}\frac{|\overline{\bf k}|}{\kappa_n}\right),
\label{eq:essential_portion}\\
& f(z) = \frac{\sin[s_0\mathrm{arcsinh}(z)]}{s_0z}\sqrt{z^2+1},
\end{align}
for $|{\bf r}|\lesssim1$ and $|\overline{\bf k}|\ll1$ (Appendix~\ref{sec:wave_function_of_Efimov_state}). 
Here the lattice spacing is adopted as the unit of length, $f(z)$ with $s_0=1.00624$ is the universal function~\cite{Gogolin:2008}, and $\kappa_n^{-1} = 1/\sqrt{m_z E_n}$ is a characteristic linear size of the $n$-th bound state:
\begin{align}
\kappa_0^{-1} = 0.799,\qquad
\kappa_1^{-1} = 14.9.
\end{align}
The wave function in Eq.~\eqref{eq:essential_portion} is normalized to satisfy $\psi_{\bf m} ({\bf r};{\bf k}_{\rm min})=f(0)=1$, where $|{\bf k}_{\rm min}|\approx 0$ is the smallest wave number.  The Efimov character of each state can be quantified by comparing the expression on the left hand side of Eq.~\eqref{eq:essential_portion}  against the universal function. The comparison shown in Fig.~\ref{fig:result2} reveals that the wave function of the $n=0$ exhibits some
deviations from the universal behavior because of the above-mentioned lattice effect. However, the excellent agreement that is obtained for the $n=1$ state at long wavelengths ($|\overline{\bf k}|<1$) confirms that these two states are indeed the bottom of the Efimov tower.

\section{Summary and discussions}\label{sec:summary_discussions}

In this paper, we predict that the magnetic field acts as a knob to tune the $s$-wave scattering length of magnons  in
Yb$_2$Ti$_2$O$_7$. 
Thus, the field plays the same role as in the Feshbach resonances of ultracold atomic gases. 
A two-magnon resonance condition is  achieved at an experimentally reachable magnetic field strength of $\sim$13 T along the [110] direction, where the scattering length diverges and the binding transition occurs. As in the case of atomic gases, Efimov states are expected to emerge near this field value. Indeed, our calculations reveal a couple of three-magnon bound states with the $s$-wave wave function  just below the three-magnon continuum of the excitation spectrum.
While the ground state ($n=0$)  exhibits some deviations from the universal character due to lattice effects,  the first excited state ($n=1$) is indeed an  Efimov state.

The results presented in this work are based on the Hamiltonian parameters of Yb$_2$Ti$_2$O$_7$ reported in Ref.~\cite{Thompson:2017}. 
Other experimental works~\cite{Ross:2011,Jaubert:2015,Bowman:2019,Robert:2015, Scheie:2019} report  larger values of $J_{zz}$, while the  other parameters are almost the same.  
The resulting critical fields for the two-magnon resonance condition are
\begin{align*}
\text{Ref.~\cite{Ross:2011} }:&~ \mu_0 H_c = 9.09~\text{T},\\ %9.088466
\text{Refs.~\cite{Jaubert:2015,Bowman:2019} }:&~ \mu_0 H_c = 8.80~\text{T},\\ %8.804617
\text{Ref.~\cite{Robert:2015} }:&~ \mu_0 H_c = 11.81~\text{T},\\ %1.1808128379379424e+01
\text{Ref.~\cite{Scheie:2019} }:&~ \mu_0 H_c = 11.48~\text{T}. % 1.1476498398282118e+01
\end{align*}
Yb$_2$Ge$_2$O$_7$ is another candidate material, whose Hamiltonian parameters have been recently reported~\cite{Sarkis:2019}. 
In this case, the two-magnon resonance is achieved at
\begin{align*}
\text{Ref.~\cite{Sarkis:2019} }:&~ \mu_0 H_c = 14.80~\text{T}.%1.4803112645303900e+01
\end{align*}
In all cases, the two-magnon resonance condition is achieved for experimentally reachable magnetic field values along the [110] direction.

Raman scattering is the ideal technique for detecting two-magnon bound states~\cite{Thorpe71,Shastry90}. 
The effective Raman operator is a linear combination of the exchange  interaction terms of the spin Hamiltonian
$\mathcal{\tilde H}_{\rm spin}$  (see Appendix B)~\cite{Shastry90}. After performing the canonical (unitary) transformation that transforms
$\mathcal{\tilde H}_{\rm spin}$ into ${\mathcal{H_{\rm eff}}}$, the resulting effective Raman operator includes terms that create pairs of bosons. These terms are responsible for the transitions between the ground state and two-magnon bound states.
Moreover, the effective Raman operator also includes terms that create three bosons, implying that three-magnon bound states can also be detected via Raman spectroscopy. These three-magnon terms of the effective Raman operator originate from
the simultaneous presence of $a^\dag_{{\bf r}}a^\dag_{{\bf r}'}$ and  $n_{\bf r} a^\dag_{{\bf r}'}$ terms in the original hard core boson model  $\mathcal{\tilde H}_{\rm spin}$.
However, the intensity of the three-magnon absorption is smaller than the one for the two-magnon bound states by a factor of order $(J/H)^2$, where $J$ represents the energy scale of $J_{z\pm}$, $J_{\pm\pm}$, or $J_\pm$. 
The binding energy of the $n=0$ Efimov state is of order $0.1$~meV [Eq.~\eqref{eq:En0}], implying that the required energy resolution is compatible with state of the art THz Raman scattering~\cite{BertoldoMenezes2018}. The same is true for the binding energy of the two-magnon bound state (green thick line in Fig.~\ref{fig:result1}). The binding energy of the $n=1$ state, on the other hand, is too small (of order $0.1$~$\mu$eV) to be detected by the existing spectroscopic techniques.
Neutron scattering can also be used to observe the two-magnon and three-magnon bound states~\cite{Garrett1997,Tennant2003}. 
However, the corresponding intensity is weak in both cases because it arises from the small hybridization of these bound states with the single-magnon state.

\begin{acknowledgments}
The authors thank Y.~ Motome and R.~Coldea for fruitful discussions.
This work was partially supported by Japan Society for the Promotion of Science (JSPS) KAKENHI Grant No. JP15K17727, No. JP15H05855, No. 16H02206 and No. 18K03447  and JST, CREST Grant No. JPMJCR18T2, Japan.
S-S Z. and C. D. B. are supported by funding from the Lincoln Chair of Excellence in Physics. Part of this work was carried out under the auspices of the U.S. DOE NNSA under contract No. 89233218CNA000001 through the LDRD Program. This research used resources of the Oak Ridge Leadership Computing Facility at the Oak Ridge National Laboratory, which is supported by the Office of Science of the U.S. Department of Energy under Contract No. DE-AC05-00OR22725.
\end{acknowledgments}

%%%%%%%%%%%%%%%%%%%%%%%
%\renewcommand{\thesection}{\Alph{section}}
%\renewcommand{\thesubsection}{\Alph{section}.\arabic{subsection}}
%\renewcommand{\thesubsubsection}{\Alph{section}.\arabic{subsection}.\arabic{subsubsection}}
%\renewcommand{\thefigure}{S\arabic{figure}}
%\renewcommand{\theequation}{S\arabic{equation}}
%\renewcommand{\thetable}{S\Roman{table}}
%\baselineskip=6mm
%%%%%%%%%%%%%%%%%%%%%%%
\appendix

\section{Spin Hamiltonian}\label{sec:spin_hamiltonian}

In this section, we specify the local spin axes and the phase factors  of the spin Hamiltonian
$\mathcal{H}_{\rm spin}$.
The spin operators are expressed in a local reference frame whose $z$-axis is parallel to the local $[111]$ direction.
Following the notation of Ref.~\cite{Thompson:2017},
the basis of the local reference frame for sublattice $\alpha=1$--$4$ reads
\begin{align}
\begin{array}{llll}
{\bf x}_1= \left(-\frac{2}{\sqrt{6}},\frac{1}{\sqrt{6}},\frac{1}{\sqrt{6}}\right),
& {\bf z}_1=\left(\frac{1}{\sqrt{3}},\frac{1}{\sqrt{3}},\frac{1}{\sqrt{3}}\right), \\
{\bf x}_2= \left(-\frac{2}{\sqrt{6}},-\frac{1}{\sqrt{6}},-\frac{1}{\sqrt{6}}\right),
& {\bf z}_2=\left(\frac{1}{\sqrt{3}},-\frac{1}{\sqrt{3}},-\frac{1}{\sqrt{3}}\right), \\
{\bf x}_3= \left(\frac{2}{\sqrt{6}},\frac{1}{\sqrt{6}},-\frac{1}{\sqrt{6}}\right),    
& {\bf z}_3=\left(-\frac{1}{\sqrt{3}},\frac{1}{\sqrt{3}},-\frac{1}{\sqrt{3}}\right), \\
{\bf x}_4= \left(\frac{2}{\sqrt{6}},-\frac{1}{\sqrt{6}},\frac{1}{\sqrt{6}}\right),
& {\bf z}_4=\left(-\frac{1}{\sqrt{3}},-\frac{1}{\sqrt{3}},\frac{1}{\sqrt{3}}\right),
\end{array}
\end{align}
and ${\bf y}_{\alpha} = {\bf z}_{\alpha} \times {\bf x}_{\alpha}$.
The $g$-tensor takes the diagonal form in the local reference frame:
\begin{align}
g = 
\begin{bmatrix}
g_\perp & 0 & 0\\
0 & g_\perp & 0\\
0 & 0 & g_\parallel
\end{bmatrix}
.
\end{align}
The phase factors in  $\mathcal{H}_{\rm spin}$ are
\begin{align}
\zeta = 
\begin{bmatrix}
 0 & -1 & e^{ {\rm i}\frac{\pi}{3}} & e^{- {\rm i}\frac{\pi}{3}}  \\
-1 & 0  & e^{-{\rm i}\frac{\pi}{3}} & e^{  {\rm i}\frac{\pi}{3}}  \\
e^{ {\rm i}\frac{\pi}{3}} & e^{- {\rm i}\frac{\pi}{3}} & 0 & -1 \\
e^{-{\rm i}\frac{\pi}{3}} & e^{  {\rm i}\frac{\pi}{3}} &-1 &  0 \\
\end{bmatrix}
,\quad
\gamma = -\zeta^*.
\end{align}

\begin{figure}[!htb]
  \centering
  \includegraphics[trim=0 0 0 0, clip,width=\columnwidth]{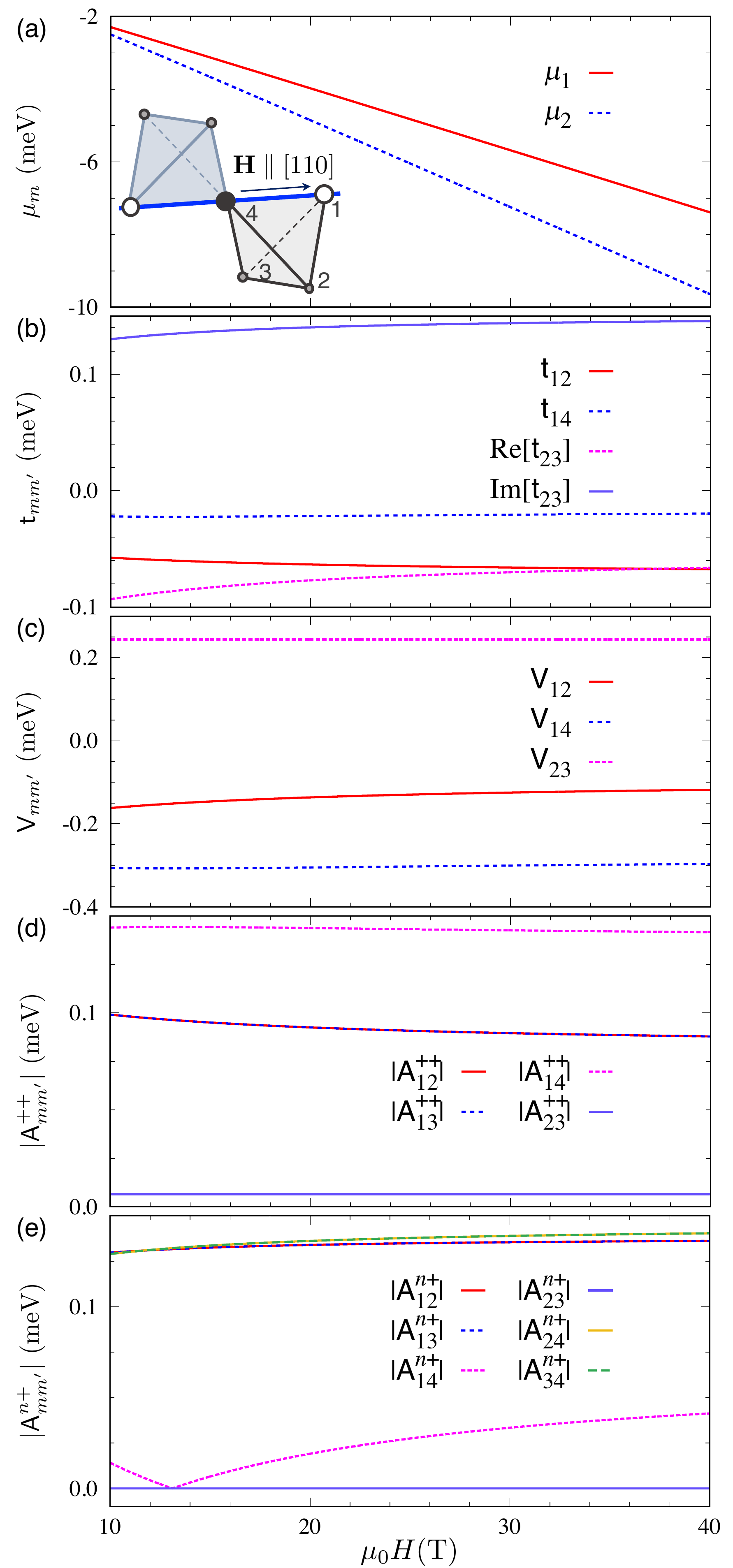}
  \caption{
  Sublattice structure of pyrochlore lattice and
  model parameters for $\mathcal{\tilde H}_{\rm spin}$.
  (a)--(e) Field dependence of the model parameters of $\mathcal{\tilde H}_{\rm spin}$  obtained by applying a local rotation to the spin Hamiltonian $\mathcal{ H}_{\rm spin}$ of Yb$_2$Ti$_2$O$_7$.
  Note that the following relation equations hold:
  $\mu_1=\mu_4$,
  $\mu_2=\mu_3$,
  $\mathsf{t}_{12}=\mathsf{t}_{13}=\mathsf{t}_{24}=\mathsf{t}_{34}$,
  $\mathsf{V}_{12}=\mathsf{V}_{13}=\mathsf{V}_{24}=\mathsf{V}_{34}$,
  $\mathsf{A}^{++}_{12}=\mathsf{A}^{++}_{24}$,
  $\mathsf{A}^{++}_{13}=\mathsf{A}^{++}_{34}$,
  $\mathsf{A}^{n+}_{12}=-\mathsf{A}^{+n}_{24}$,
  $\mathsf{A}^{n+}_{13}=-\mathsf{A}^{+n}_{34}$,
  $\mathsf{A}^{n+}_{14}=-\mathsf{A}^{+n}_{14}$,
  $\mathsf{A}^{n+}_{24}=-\mathsf{A}^{+n}_{12}$,
  and
  $\mathsf{A}^{n+}_{34}=-\mathsf{A}^{+n}_{13}$. 
  Inset of (a) shows schematic view of the pyrochlore lattice with sublattice indices 1--4.
  }
  \label{fig:org_mdl_param}
\end{figure}

%\clearpage

\section{Hard-core boson representation of the spin Hamiltonian}\label{sec:hardcoreboson_hamiltonian}

In this section, we express the spin Hamiltonian in a new local reference frame and then apply a Matsubara-Matsuda transformation~\cite{Matsubara:1956}
(exact mapping between spin 1/2 operators and hard-core bosons).
The new local reference frame, defined by the three axes $(\tilde{{\bf x}}_{\alpha}, \tilde{{\bf y}}_{\alpha},\tilde{{\bf z}}_{\alpha})$, is such that $\tilde{{\bf z}}_{\alpha}$ is parallel to the direction of local magnetic moments, ${\boldsymbol{\mathsf m}}_{\alpha}$,
that minimizes the classical limit of $\mathcal{H}_{\rm spin}$. In other words,  $\tilde{{\bf z}}_{\alpha}$ is parallel to the direction of the magnetic moment that is obtained from a mean field decoupling of the exchange interaction in $\mathcal{H}_{\rm spin}$.
In the new reference frame, we map the spin 1/2 operators into creation and annihilation operators of hard-core bosons:
\begin{align}
& \tilde{{\mathsf{S}}}^z_{\bf r} = \frac{1}{2} -  n_{\bf r} ,\\
& \tilde{\mathsf{S}}^{+}_{\bf r} = \tilde{\mathsf{S}}_{{\bf r}}  ^x + {\rm i} \tilde{\mathsf{S}}_{{\bf r}}  ^y = a^{\;}_{{\bf r}}  ,\\
& \tilde{\mathsf{S}}^{-}_{\bf r} = \tilde{\mathsf{S}}_{{\bf r}}  ^x - {\rm i} \tilde{\mathsf{S}}_{{\bf r}}  ^y = a^{\dag}_{{\bf r}}  ,
\end{align}
with $n_{\bf r} = a^\dag_{\bf r} a^{\;}_{{\bf r}}$. 
The hard-core condition, $(a_{\bf r}^{\dagger})^2 = 0, \forall {\bf r}$, is necessary to keep the  dimension of the local Hilbert space equal to
$2$. The original spin Hamiltonian [Eq.~\eqref{eq:Hspin} in the main text] can then be reexpressed as a  Hamiltonian for a gas of hard-core bosons, whose particle number is not conserved. Up to an irrelevant constant, we obtain
\begin{align}
\mathcal{\tilde H}_{\rm spin} =&
\sum_{\langle {\bf rr}' \rangle}
\big[
(\mathsf{t}_{\alpha_{\bf r} \alpha_{{\bf r}'}  }
a^\dag_{\bf r} a^{\;}_{{\bf r}'}  
+{\rm h.c.}) 
+
\mathsf{V}_{\alpha_{\bf r} \alpha_{{\bf r}'}  } n_{\bf r} n_{{\bf r}'}  \nonumber \\
&\quad+
(
\mathsf{A}^{++}_{\alpha_{\bf r} \alpha_{{\bf r}'}  }
 a^\dag_{\bf r} a^\dag_{{\bf r}'}  
+{\rm h.c.}
) \nonumber\\
&\quad+
(
\mathsf{A}^{n+}_{\alpha_{\bf r} \alpha_{{\bf r}'}  }
n^{\;}_{\bf r} a^\dag_{{\bf r}'}  
+
\mathsf{A}^{+n}_{\alpha_{\bf r} \alpha_{{\bf r}'}  }
a^\dag_{\bf r} n^{\;}_{{\bf r}'}  
+{\rm h.c.}
)
\big] 
\nonumber \\
&\quad
-\sum_{\bf r} \mu_{\alpha_{{\bf r}}  } n_{{\bf r}}  
+ \mathsf{U} \sum_{\bf r} n_{\bf r} (n_{{\bf r}}  -1).
\label{eq:org_mdl}
\end{align}
The on-site repulsion $\mathsf{U}=\infty$ enforces the hard-core constraint, while the other model parameters $\mu_\alpha$, $\mathsf{t}_{\alpha \alpha'}$,
$\mathsf{V}_{\alpha \alpha'}$, $\mathsf{A}^{++}_{\alpha \alpha'}$, $\mathsf{A}^{n+}_{\alpha \alpha'}$, and $\mathsf{A}^{+n}_{\alpha \alpha'}$ depend on the external field $H$. 
We note that the choice $\tilde{{\bf z}}_{\alpha} \parallel \boldsymbol{\mathsf{m}}_{\alpha}$ eliminates all the linear 
 terms in $a^{\;}_{{\bf r}}  $ and  $a_{{\bf r}}  ^\dag$ from $\mathcal{\tilde H}_{\rm spin}$.

Figure~\ref{fig:org_mdl_param} shows the field dependence of the parameters of $\mathcal{\tilde H}_{\rm spin}$
that are obtained for the exchange matrix and the $g$-tensor that have been reported for Yb$_2$Ti$_2$O$_7$~\cite{Thompson:2017}. 
Through a proper $U(1)$ gauge transformation, $a_{\bf r} \rightarrow a_{\bf r} e^{i\theta_{\bf r}}$,  it is possible to make all the parameters $\mathsf{t}_{\alpha_{\bf r} \alpha_{{\bf r}'}  }$ real, except for $\mathsf{t}_{23}$, and, in addition, $\mathsf{t}_{12}=\mathsf{t}_{42}=\mathsf{t}_{31}=\mathsf{t}_{34}  < 0$ and $\mathsf{t}_{14} < 0$.

\section{Derivation of the effective boson model}\label{sec:derivation_Heff}

\begin{figure}[t]
  \centering
 \includegraphics[trim=0 0 0 0, clip,width=\columnwidth]{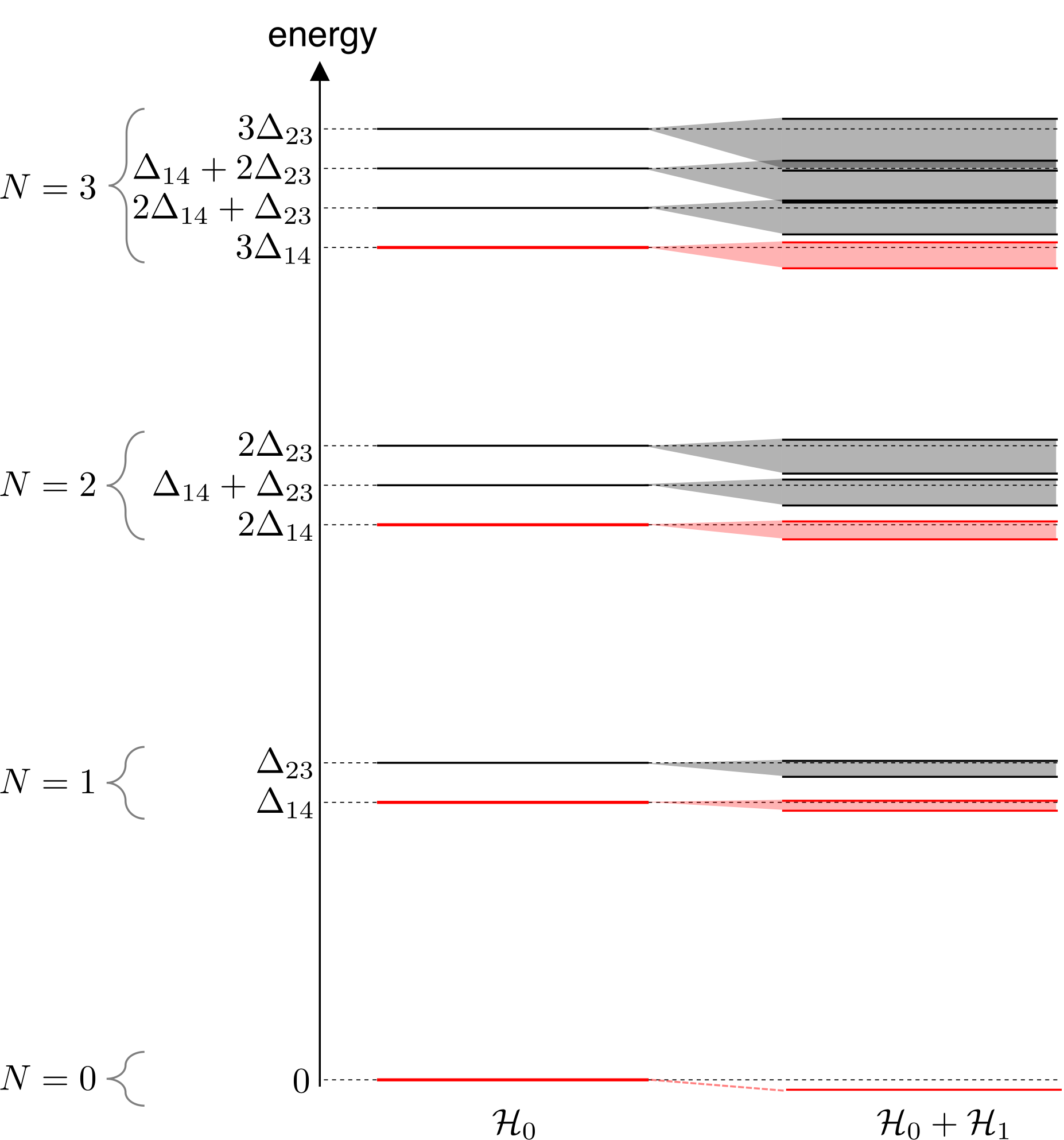}
  \caption{
  Schematic view of energy spectrum of the unperturbed Hamiltonian $\mathcal{H}_0$ and the full Hamiltonian $\mathcal{H}_0 + \mathcal{H}_1$ in the $N$-magnon sector with $N=0\text{--}3$.
  The red levels correspond to the states where all magnons occupy the low-energy sublattices 1 or 4.
  }
  \label{fig:Espe_perturb}
\end{figure}

The effective Hamiltonian, $\mathcal{H}_{\rm eff}$, is derived from $\mathcal{\tilde H}_{\rm spin}$ in the high-field regime by applying the second order perturbation theory.
For a strong enough field $H$, the chemical potential term becomes the dominant energy scale.
We then divide $\mathcal{\tilde H}_{\rm spin}$ into two parts:
\begin{align}
\mathcal{\tilde H}_{\rm spin} =&~ \mathcal{H}_0 + \mathcal{H}_1,\\
\mathcal{H}_0 =&
-\sum_{\bf r} \mu_{\alpha_{{\bf r}}  } n_{{\bf r}}  
+ \mathsf{U} \sum_{\bf r} n_{\bf r} (n_{{\bf r}}  -1),\\
\mathcal{H}_1 =&
\sum_{\langle {\bf rr}' \rangle}
\big[
(\mathsf{t}_{\alpha_{\bf r} \alpha_{{\bf r}'}  }
a^\dag_{\bf r} a^{\;}_{{\bf r}'}  
+{\rm h.c.}) +
\mathsf{V}_{\alpha_{\bf r} \alpha_{{\bf r}'}  } n_{\bf r} n_{{\bf r}'}   \nonumber\\
&\quad+
(
\mathsf{A}^{++}_{\alpha_{\bf r} \alpha_{{\bf r}'}  }
 a^\dag_{\bf r} a^\dag_{{\bf r}'}  
+{\rm h.c.}
) \nonumber\\
&\quad+
(
\mathsf{A}^{n+}_{\alpha_{\bf r} \alpha_{{\bf r}'}  }
n^{\;}_{\bf r} a^\dag_{{\bf r}'}  
+
\mathsf{A}^{+n}_{\alpha_{\bf r} \alpha_{{\bf r}'}  }
a^\dag_{\bf r} n^{\;}_{{\bf r}'}  
+{\rm h.c.}
)
\big],
\label{eq:org_mdl2}
\end{align}
where $\mathcal{H}_0$ is the unperturbed part and $\mathcal{H}_1$ is the perturbation.
The energy spectrum of the unperturbed Hamiltonian has discrete energy levels $E(N_{14}, E_{23}) = \Delta_{14} N_{14} + \Delta_{23} N_{23} $, where $\Delta_{\alpha} \equiv -\mu_\alpha$, $\Delta_{14}=\Delta_1 = \Delta_4$, $\Delta_{23}=\Delta_2 = \Delta_3$, and $N_{14}$ ($N_{23}$) is the number of hard-core bosons in sublattices 1 and 4 (2 and 3)~(Fig.~\ref{fig:Espe_perturb}).
$\mathcal{H}_0$ is massively degenerate in each sector $(N_{14}, N_{23})$ except for $(N_{14}, N_{23})=(0,0)$.
The eigenvalues for different sectors are separated by an energy gap proportional to the field strength $H$.
In the high-field regime, one can construct an effective Hamiltonian acting on each sector by treating $\mathcal{H}_1$ as a perturbation.
For each degenerate subspace $\mathcal{S}_E$ of $\mathcal{H}_0$ with eigenenergy $E$,
we introduce a projector $\mathcal{P}_E$ and the orthogonal projector $\mathcal{Q}_E = 1 -\mathcal{P}_E$.
To second order in the perturbation,  the effective Hamiltonian acting on $\mathcal{S}_E$ is given by
\begin{align}
\mathcal{H}_{\rm eff} (E) =&~
E+
\mathcal{P}_E \mathcal{H}_1 \mathcal{P}_E
+
\mathcal{P}_E
\mathcal{H}_1
\mathcal{Q}_E
\frac{1}{E-\mathcal{H}_0}
\mathcal{Q}_E
\mathcal{H}_1
\mathcal{P}_E.
\label{eq:2nd_pert}
\end{align}
We are interested in the effective low-energy Hamiltonian that is obtained by projecting on the lowest energy sector for each total number of magnons $N=N_{14}+N_{23}$. This is simply the sector that satisfies $N=N_{14}$ and $N_{23}=0$.
For instance, let us consider the single-particle hopping amplitude $t_\parallel$ between nearest-neighbor sites of the low-energy chains 1 or 4~[Inset of Fig.~\ref{fig:org_mdl_param}(a)].
While the second term on the right hand side of Eq.~\eqref{eq:2nd_pert} is simply $\mathsf{t}_{14} a^\dag_4 a^{\;}_1$, the third term has contributions from multiple perturbation processes,
including a number-conserving perturbation process where a magnon tunnels via  sublattice 2 or sublattice 3, e.g., hopping process $1\to2\to4$.
By taking into account all the other perturbation processes, we obtain
the nearest-neighbor intra-chain  hopping amplitude,
\begin{align}
t_{\parallel} =&
{\mathsf{t}}_{14}
-\frac{ 
 {\mathsf{A}}^{++}_{12} ( {\mathsf{A}}^{++}_{24})^*
+ {\mathsf{A}}^{++}_{13} ( {\mathsf{A}}^{++}_{34})^*
}{\Delta_{14}+\Delta_{23}} \nonumber\\
& -\frac{ 
 {\mathsf{A}}^{n+}_{14} ( {\mathsf{A}}^{+n}_{14})^*
}{\Delta_{14}}
-\frac{ 
 {\mathsf{t}}_{12}  {\mathsf{t}}_{24}
+ {\mathsf{t}}_{13}  {\mathsf{t}}_{34}
}{\Delta_{23}-\Delta_{14}}
.\label{eq:tpara_pert}
\end{align}

\begin{figure}[t]
  \centering
 \includegraphics[trim=0 0 0 0, clip,width=\columnwidth]{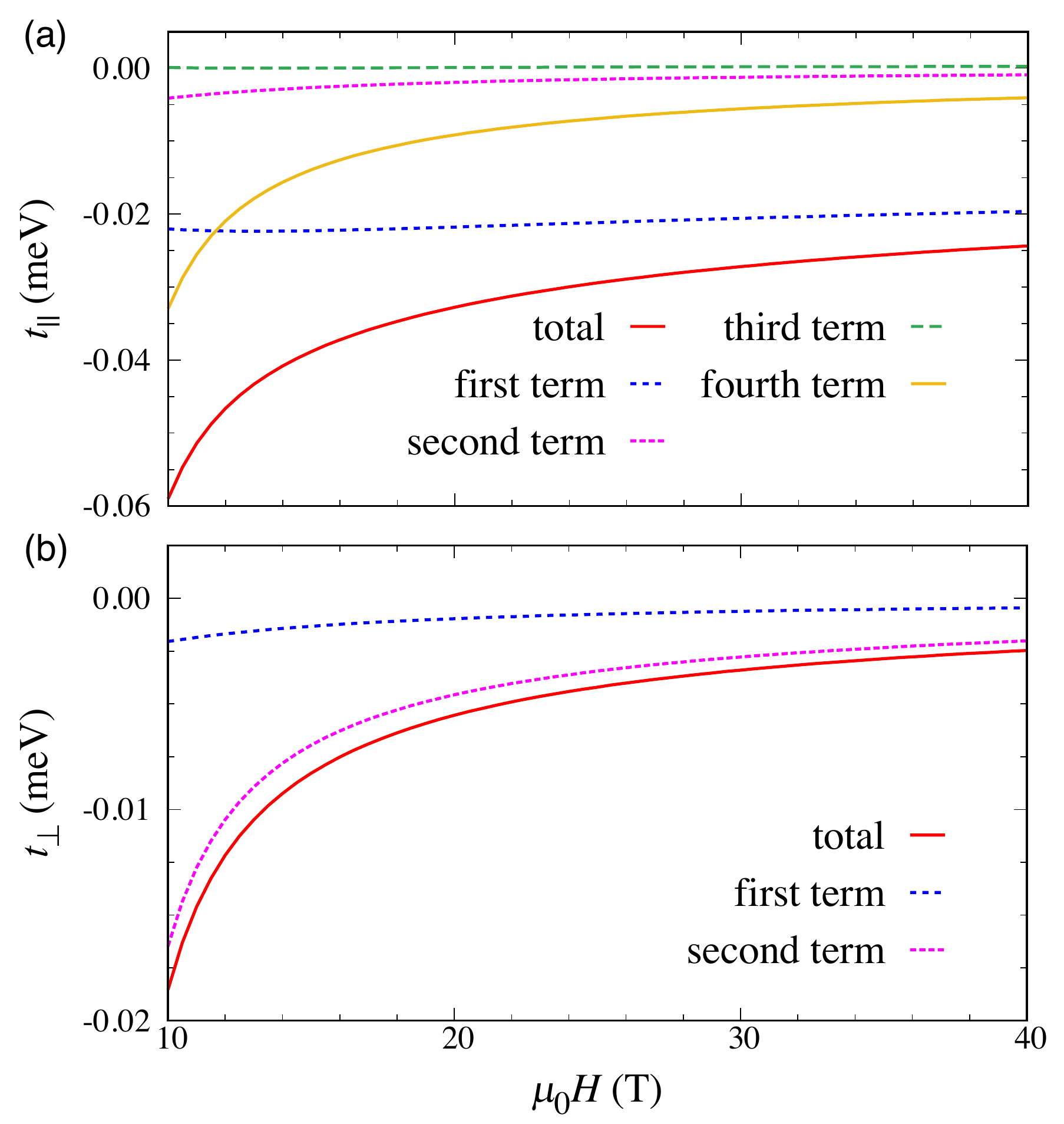}
  \caption{
	Decomposition of the hopping amplitudes.
	(a) $t_\parallel$ and (b) $t_\perp$ are decomposed into
	four and two terms as in the right hand sides of Eqs.~\eqref{eq:tpara_pert} and \eqref{eq:tperp_pert}, respectively.
  }
  \label{fig:tdecomposition}
\end{figure}

The other model parameters are obtained in a similar way,
\begin{align}
t'_{\parallel} =&
-\frac{ 
| {\mathsf{A}}^{++}_{14} |^2
}{2\Delta_{14}},\\
t_{\perp}=t'_{\perp}=&
-\frac{ 
 {\mathsf{A}}^{++}_{12} ( {\mathsf{A}}^{++}_{24})^*
}{\Delta_{23}+\Delta_{14}} 
-\frac{ 
 {\mathsf{t}}_{12}  {\mathsf{t}}_{24}
}{\Delta_{23}-\Delta_{14}} \nonumber\\
=&
-\frac{ 
 {\mathsf{A}}^{++}_{13} ( {\mathsf{A}}^{++}_{34})^*
}{\Delta_{23}+\Delta_{14}} 
-\frac{ 
 {\mathsf{t}}_{13}  {\mathsf{t}}_{34}
}{\Delta_{23}-\Delta_{14}}
,\label{eq:tperp_pert}\\
\mu =&
\mu_1
-2\sum_{\gamma=2,3,4} \frac{|  {\mathsf{A}}^{++}_{1\gamma} |^2 }{\Delta_1 + \Delta_\gamma}
+2\sum_{\gamma=2,3}   \frac{ {\mathsf{t}}_{1\gamma} {\mathsf{t}}_{\gamma1}  }{\Delta_\gamma - \Delta_1} \nonumber\\
&+2\sum_{\gamma=2,3,4} \frac{ {\mathsf{A}}^{n+}_{1\gamma} ( {\mathsf{A}}^{n+}_{1\gamma})^*  }{\Delta_\gamma },\\
u_\parallel =&
\mathsf{V}_{14}
+ 
\left[ -\sum_{\gamma=2,3} \frac{
( {\mathsf{A}}^{n+}_{1\gamma})^*  {\mathsf{A}}^{+n}_{\gamma 4}
+( {\mathsf{A}}^{n+}_{4\gamma})^*  {\mathsf{A}}^{+n}_{\gamma 1}
}{\Delta_\gamma}
\right] \nonumber\\
&+2\frac{ | {\mathsf{A}}^{n+}_{41} |^2 + | {\mathsf{A}}^{n+}_{14} |^2 }{\Delta_{14}}
,\\
u'_\parallel =&~ -2 \frac{ ( {\mathsf{A}}^{n+}_{14})^*  {\mathsf{A}}^{+n}_{41} }{\Delta_{14}},\\
u_\perp =&~ -\frac{ ( {\mathsf{A}}^{n+}_{12})^*  {\mathsf{A}}^{+n}_{24} +  {\mathsf{A}}^{n+}_{12}( {\mathsf{A}}^{+n}_{24})^*  }{\Delta_{23}},\\
v_1 =&~ -\frac{| {\mathsf{A}}^{++}_{14}|^2}{-2\Delta_{14}} - t'_{\parallel},\\
v_2 =&~ -\frac{( {\mathsf{A}}^{n+}_{14})^*  {\mathsf{A}}^{n+}_{41} }{\Delta_{14}},\\
w=&~2\frac{
|
{\mathsf A}_{14}^{+n}
|^2}{
\Delta_{14}}
-u'_\parallel.
\end{align}
The effective hopping amplitudes $t_\parallel$ and $t_\perp$ have strong field dependence because of the dominant tunneling processes via the high-energy chains 2 and 3.
Figure~\ref{fig:tdecomposition} shows the field dependence of  each term on the right hand sides of Eqs.~\eqref{eq:tpara_pert} and \eqref{eq:tperp_pert}, as well as the total values.
The fourth term of Eq.~\eqref{eq:tpara_pert}
and the second term of Eq.~\eqref{eq:tperp_pert} give the dominant contributions to the field dependence of the hopping amplitude. 
These terms correspond to the number-conserving perturbation processes discussed above.

In the main text, we focus on the resonance condition $\mu_0 H = 12.91$ T where the model parameters  (in meV) are:
\begin{align*}
t_\parallel =& -0.04357,\\
t'_{\parallel} =& -0.003748,\\
t_{\perp}=t'_{\perp}=& -0.01061,\\
\mu =& -2.736,\\
u_\parallel =& -0.2856,\\
u'_\parallel =& -4.191 \times 10^{-7},\\
u_\perp =& 0.01090.\\
v_1 =& 0.007495,\\
v_2 =& 2.095 \times 10^{-7},\\
w=&8.382 \times 10^{-7}.
\end{align*}

\section{Solution of the few-body problem}\label{sec:solution_of_few_body_problem}

\subsection{Convenient notation of lattice site coordinates}
We first introduce a new notation for the lattice sites, which turns out to be convenient for implementing
symmetry operations on the Lippmann-Schwinger equation.  These operations are used to reduce the computational cost of solving the three-magnon problem. 
Each lattice site is specified by ${\bf r}=r_x {\bf e}_x + r_y {\bf e}_y + r_z {\bf e}_z$
where $r_{x,y}=1,2, \cdots , L$, $r_z = 1,2, \cdots , 2L$,
with ${\bf e}_x = {\bf a}_1- {\bf a}_3/2$, ${\bf e}_y = {\bf a}_2 - {\bf a}_3/2$, and ${\bf e}_z = {\bf a}_3/2$.
The sign $ m_{\bf r} \equiv (-1)^{r_x+r_y+r_z}$ is the sublattice index,
namely $m_{\bf r}=+1$ ($-1$) corresponds to the sublattice $\alpha_{\bf r} = 4$ ($1$)~(Fig.~\ref{fig:lattice}). 
The primitive translation vectors (${\bf a}_{1,2,3}$) and the reciprocal vectors (${\bf G}_{1,2,3}$)
of the pyrochlore lattice are
\begin{align}
\begin{cases}
{\bf a}_1 = \sqrt{2}\ell (0,1,1)\\
{\bf a}_2 = \sqrt{2}\ell (1,0,1)\\
{\bf a}_3 = \sqrt{2}\ell (1,1,0)\\
\end{cases},\quad\quad
\begin{cases}
{\bf G}_1 = \frac{\pi}{\sqrt{2} \ell}  (-1,1,1)\\
{\bf G}_2 = \frac{\pi}{\sqrt{2} \ell}  (1,-1,1)\\
{\bf G}_3 = \frac{\pi}{\sqrt{2} \ell}  (1,1,-1)\\
\end{cases},
\end{align}
where $\ell$ is the distance between nearest-neighbor Yb cation pairs.
We take $\ell$ as the unit of length. For a finite lattice of $(2\times L^3)$-sites  with even $L$,
the wave vectors in the first Brillouin zone are given by
\begin{align}
{\bf k} = \sum_{d=1}^3 {\bf G}_d \frac{n_d}{L},~~
\sbra{
	n_d=-\frac{L}2,\cdots,\frac{L}{2}-1
	}.
\end{align}
The summation over ${\bf k}$ becomes an integral in the infinite volume limit $L \to \infty$:
\begin{align}
\frac{1}{L^3} \sum_{\bf k} \to
 \int_{\bf k} \equiv& \iiint_{-\pi}^\pi\frac{dk_1 dk_2 dk_3 }{(2\pi)^3}\Bigg|_{
 \displaystyle 
{\bf k}=\frac{1}{2\pi} \sum_{d=1}^3 k_d {\bf G}_d
}.
\end{align} 
The integration over the first Brillouin zone is redefined by shifting the wave vector ${\bf k}$:
\begin{align}
&\begin{cases}
\tilde{k}_x = k_1 -\frac{k_3}2\\
\tilde{k}_y = k_2 -\frac{k_3}2\\
\tilde{k}_z = k_3\\
\end{cases}
\quad: \nonumber\\
&\int_{\bf k}
\to \iiint_{-\pi}^\pi\frac{d\tilde{k}_xd \tilde{k}_yd \tilde{k}_z}{(2\pi)^3}\Bigg|_{
%\displaystyle
{\bf k}=
\frac{1}{2\pi}
\sbra{
  \tilde{k}_x  \tilde{\bf G}_x
+ \tilde{k}_y  \tilde{\bf G}_y
+ \frac{\tilde{k}_z}{2} \tilde{\bf G}_z
}
},
\end{align}
with
\begin{align}
\tilde{\bf G}_x = {\bf G}_1,\quad
\tilde{\bf G}_y = {\bf G}_2,\quad
\tilde{\bf G}_z = {\bf G}_1 + {\bf G}_2 + 2 {\bf G}_3 .
\end{align}
We choose the subscript $\{ x,y,z \}$ for $\tilde{\bf G}$  instead of $\{1,2,3\}$ because
$\tilde{\bf G}_{x,y,z}$ can be regarded as the reciprocal  vectors of the lattice spanned by 
the primitive vectors ${\bf e}_{x,y,z}$,
\begin{align}
2\pi \frac{ \tilde{\bf G}_\mu \times \tilde{\bf G}_\nu }{
\tilde{\bf G}_x \cdot ( \tilde{\bf G}_y \times \tilde{\bf G}_z )
}=\sum_{\gamma}\epsilon_{\mu\nu\gamma} {\bf e}_\gamma,
\end{align}
where $\epsilon_{\mu\nu\gamma}$ is the Levi-Civita tensor.

In this notation, the effective boson Hamiltonian becomes
\begin{align}
\mathcal{H}_{\rm eff}
=&
\sum_{\bf r} \Big[ \sum_{\bf e} t^{m_{\bf r}}_{\bf e} a^\dag_{\bf r} a^{\;}_{{\bf r}+{\bf e}}
+ \sum_{{\bf e},{\bf e}'} u^{m_{\bf r}m_{{\bf r}+{\bf e}}}_{{\bf e}{\bf e}'} a^\dag_{\bf r} a^\dag_{{\bf r}+{\bf e}} a^{\;}_{{\bf r}+{\bf e}'} a^{\;}_{{\bf r}}
\Big] \nonumber\\
&-\mu \sum_{\bf r} 
a^\dag_{\bf r} a^{\;}_{{\bf r}}.
\end{align}
The hopping amplitudes $t^m_{\bf e}$ and the two-body interactions $u^{mm'}_{{\bf ee}'}$
are defined as
\begin{align}
t^{m}_{\bf e} =&
\begin{cases}
t_\parallel &\text{if } {\bf e}=\pm {\bf e}_z;\\
t'_\parallel &\text{if } {\bf e}=\pm 2 {\bf e}_z;\\
t_\perp &\text{if }  {\bf e}=\pm {\bf e}_{x,y}, \pm ({\bf e}_{x,y} + m {\bf e}_z );\\
0 &{\rm otherwise},
\end{cases}\\
u^{mm'}_{{\bf ee}'} =&
\begin{cases}
 u_\parallel &\text{if } {\bf e}={\bf e}'={\bf e}_z,~m=\overline{m'}; \\
 u_\perp     &\text{if } {\bf e}={\bf e}'={\bf e}_{x,y},~m=\overline{m'};\\
-u_\perp     &\text{if }  {\bf e}={\bf e}'= ({\bf e}_{x,y} + m {\bf e}_z ),~m=m';\\
 v_1 & \text{if }  {\bf e}= -{\bf e}' = \pm {\bf e}_z,~m=\overline{m'};\\
 \mathsf{U}=\infty& \text{if } {\bf e}={\bf e}' = 0,~m=m';\\
0 &{\rm otherwise},
\end{cases}
\end{align}
where $\overline{m} \equiv -m$.
Note that we neglected $u_\parallel'$, $v_2$, and $w$ because they are much smaller than the other interactions as discussed in the main text (Sec.~\ref{sec:model}).

In what follows, we consider the one-, two-, and three-magnon subspaces spanned by
one-magnon states ${a}^\dag_{{\bf r}_1} \left| \emptyset \right\rangle$,
two-magnon states ${a}^\dag_{{\bf r}_1} {a}^\dag_{{\bf r}_2} \left| \emptyset \right\rangle$,
and three-magnon states ${a}^\dag_{{\bf r}_1} {a}^\dag_{{\bf r}_2} {a}^\dag_{{\bf r}_3} \left| \emptyset \right\rangle$,
respectively. The boson vacuum $\left| \emptyset \right\rangle$ represents the ground state.

\subsection{Single-magnon problem}\label{sec:single_magnon_problem}

The projection of the Schr\"odinger equation, $E\ket{\Psi}= \mathcal{H}_{\rm eff} \ket{\Psi}$, 
onto the  single-magnon basis states $a_{\bf r}^\dag \ket{\emptyset}$ leads to
\begin{align}
& E \Psi({\bf r}) = \bracketii{\emptyset}{ a_{\bf r} \mathcal{H}_{\rm eff} }{\Psi} 
= \sum_{\bf e} t_{\bf e}^{m_{\bf r}}\Psi({\bf r}+{\bf e}),
\end{align}
where $\ket{\Psi}$ is in the one-magnon subspace,  and $\Psi({\bf r}) = \bracketii{\emptyset}{ a_{\bf r} }{\Psi}$.
From the Fourier transform, 
\begin{align}
\tilde\Psi_m({\bf k}) = \sum_{{\bf r} \in m}e^{-{\rm i} {\bf k} \cdot {\bf r} } \Psi({\bf r}),~
\Psi({\bf r}) = \frac1{L^3} \sum_{\bf k} e^{{\rm i} {\bf k} \cdot {\bf r}} \tilde\Psi_{m_{\bf r}} ({\bf k}),
\end{align}
where the sum of ${\bf r} \in m$ runs over all the lattice sites of a given sublattice, $m_{\bf r} = m$,
we obtain
\begin{align}
E
\begin{bmatrix}
\tilde\Psi_+({\bf k}) \\ \tilde\Psi_-({\bf k})
\end{bmatrix}
=
\begin{bmatrix}
\varepsilon_{{\bf k}++} & \varepsilon_{{\bf k}+-} \\
\varepsilon_{{\bf k}-+} & \varepsilon_{{\bf k}--} \\
\end{bmatrix}
\begin{bmatrix}
\tilde\Psi_+({\bf k}) \\ \tilde\Psi_-({\bf k})
\end{bmatrix},
\end{align}
where
\begin{align}
&\varepsilon_{{\bf k}mm}
=
-\mu+
2t'_\parallel\cos(2{\bf k} \cdot {\bf e}_z) \nonumber\\
&\quad\quad+ 2t_\perp [\cos({\bf k}\cdot {\bf e}_x + m {\bf k}\cdot {\bf e}_z) + \cos({\bf k}\cdot {\bf e}_y + m {\bf k}\cdot {\bf e}_z)  ]
,\nonumber\\
&\varepsilon_{{\bf k}m \overline{m}} =
2t_\parallel \cos({\bf k} \cdot  {\bf e}_z) + 2t_\perp [\cos({\bf k}\cdot {\bf e}_x)+ \cos({\bf k}\cdot {\bf e}_y)].
%\begin{bmatrix}
%2t'_\para\cos(2\k\cdot\hat\e_z) + 2t_\perp\sum_{\mu=x,y}\cos(\k\cdot(\hat\e_\mu+\hat\e_z))
%& 2t_\para\cos(\k\cdot\hat\e_z) + 2t_\perp\sum_{\mu=x,y}\cos(\k\cdot\hat\e_\mu) \\
%2t_\para\cos(\k\cdot\hat\e_z) + 2t_\perp\sum_{\mu=x,y}\cos(\k\cdot\hat\e_\mu)
%& 2t'_\para\cos(2\k\cdot\hat\e_z) + 2t_\perp\sum_{\mu=x,y}\cos(\k\cdot(\hat\e_\mu-\hat\e_z))
%\end{bmatrix}
\end{align}
The single-magnon spectrum is given by the eigenvalues of the matrix $\varepsilon_{\bf k}$.

\subsection{Two-magnon problem}\label{sec:two_magnon_problem}
In this subsection, we explain how to compute the $s$-wave scattering length and the binding energy of the two-magnon bound state using the Lippmann-Schwinger equation.
Similarly to the single-magnon problem, the projection of the Schr\"odinger equation, 
$E\ket{\Psi}=
\mathcal{H}_{\rm eff}
\ket{\Psi}$, onto the (unnormalized) two-magnon basis states $a_{{\bf r}_1}^\dag a_{{\bf r}_2}^\dag \ket{\emptyset} $  leads to
\begin{align}
E \Psi ({\bf r}_1, {\bf r}_2) =&
\bracketii{\emptyset}{ a_{{\bf r}_2} a_{{\bf r}_1} 
\mathcal{H}_{\rm eff}
}{\Psi} \nonumber \\
=& \sum_{\bf e}
\bbra{ t_{\bf e}^{m_{{\bf r}_1}}\Psi({\bf r}_1+{\bf e},{\bf r}_2)
+  t_{\bf e}^{m_{{\bf r}_2}}\Psi({\bf r}_1,{\bf r}_2+{\bf e}) } \nonumber \\
&+ \sum_{{\bf e}, {\bf e}' } 
\big[
\delta_{{\bf r}_2 \; {\bf r}_1+{\bf e}}
u_{{\bf ee}'}^{m_{{\bf r}_1}m_{{\bf r}_1+{\bf e}}}\Psi({\bf r}_1,{\bf r}_1+{\bf e}') \nonumber\\
&\quad\quad+ \delta_{{\bf r}_1 \; {\bf r}_2+{\bf e}}
u_{{\bf ee}'}^{m_{{\bf r}_2}m_{{\bf r}_2+{\bf e}}}\Psi({\bf r}_2+{\bf e}',{\bf r}_2)
\big],
\end{align}
where $\Psi({\bf r}_1,{\bf r}_2)=\bracketii{\emptyset}{a_{{\bf r}_2} a_{{\bf r}_1} }{\Psi}$.
Here, we focus on the two-magnon scattering in the long-wavelength limit (${\bf k} \to 0$) to compute the $s$-wave scattering length $a$.
For this purpose, we consider the bottom of the two-magnon continuum with zero center-of-mass momentum ${\bf K}={\bf 0}$ and energy $E = 2 \mathcal{E}_0$ ($\mathcal{E}_0=-\mu+ 2t_\parallel+2t'_\parallel+8 t_\perp$). 
From the Fourier transform,
\begin{align}
\tilde\Psi_{m^{\;}_1m^{\;}_2}({\bf k}_1,{\bf k}_2) =& 
\sum_{{\bf r}_1 \in m^{\;}_1,{\bf r}_2 \in m^{\;}_2}
e^{-{\rm i} \sbra{ {\bf k}_1\cdot {\bf r}_1+ {\bf k}_2\cdot{\bf r}_2 } }\Psi( {\bf r}_1,{\bf r}_2), \nonumber%\\
\end{align}\begin{align}
\Psi({\bf r}_1,{\bf r}_2) =&  
\frac1{L^6}\sum_{{\bf k}_1,{\bf k}_2}
e^{{\rm i} \sbra{ {\bf k}_1\cdot {\bf r}_1+ {\bf k}_2 \cdot {\bf r}_2} }  \tilde\Psi_{ m_{{\bf r}_1}m_{{\bf r}_2} } ({\bf k}_1,{\bf k}_2),
\end{align}
we obtain
\begin{widetext}
\begin{align}
& \sum_{m'_1,m'_2}\bbra{
E\delta_{m^{\;}_1 m'_1}\delta_{m^{\;}_2 m'_2}
-\varepsilon_{{\bf k}_1 m^{\;}_1 m_1'}\delta_{m^{\;}_2m_2'}
-\varepsilon_{{\bf k}_2 m^{\;}_2 m_2'}\delta_{m^{\;}_1m_1'}
}
\sbra{
\tilde\Psi_{m'_1m'_2}({\bf k}_1,{\bf k}_2) 
-\delta_{{\bf k}_1{\bf 0}} \delta_{{\bf k}_2 {\bf 0}} \phi_{m_1' m_2'} L^3
}\nonumber\\
&= \frac1{L^3}\sum_{{\bf k}'_2} \sum_{{\bf e},{\bf e}'}
\left[e^{-{\rm i} \sbra{ {\bf k}_2\cdot {\bf e}- {\bf k}'_2\cdot {\bf e}' }}u_{{\bf ee}'}^{m_1m_2}
+ e^{-{\rm i} \mbra{ {\bf k}_1\cdot {\bf e} - \sbra{ {\bf k}_1+{\bf k}_2-{\bf k}'_2 }\cdot {\bf e}' } }u_{{\bf ee}'}^{m_2m_1}\right]
\tilde\Psi_{m^{\;}_1m^{\;}_2}({\bf k}_1+{\bf k}_2-{\bf k}'_2,{\bf k}'_2),
\label{eq:Sch_eq_2body}
\end{align}
\end{widetext}
where 
$m_{\bf e}=m_{{\bf e}'}$ is always satisfied for finite $u_{{\bf ee}'}^{mm'}$. 
$\phi_{m'_1 m'_2}$ represents the two-magnon eigenstate of  the non-interacting problem with eigenvalue $2\mathcal{E}_0$ in momentum space,
\begin{align}
&\sum_{m'_1, m'_2}
\bbra{ 
	\varepsilon_{{\bf 0} m_1 m_1'} \delta_{m_2 m'_2} 
	+\varepsilon_{{\bf 0} m_2 m_2'} \delta_{m_1 m'_1} 
	}\phi_{m'_1 m'_2} = 2 \mathcal{E}_0 \phi_{m_1 m_2},\\
&\sum_{m'_1, m'_2}	
\phi^*_{m'_1 m'_2}
\phi^{\;}_{m'_1 m'_2} =1,
\end{align}
that has a simple solution, $\phi_{m_1m_2} = 1/2 \; \forall (m_1,m_2)$, for  ${\bf k}_1={\bf k}_2=0$. % and $E = 2 \mathcal{E}_0$.
We only consider states with zero center-of-mass momentum and impose the ansatz 
$\tilde\Psi_{mm'}({\bf k}_1,{\bf k}_2) = \delta_{{\bf k}_1\;-{\bf k}_2}\tilde\psi_{mm'}({\bf k}_2)$, which is symmetric under an exchange of two bosons: $\tilde\psi_{mm'}({\bf k}) = \tilde\psi_{m'm}(-{\bf k})$. The unknown functions $\tilde\psi_{mm'}({\bf k})$ satisfy the Lippmann-Schwinger equation:
\begin{widetext}
\begin{align}
\tilde\psi_{m^{\;}_1m^{\;}_2}({\bf k})
&= 
\delta_{{\bf k0}} \phi_{m_1m_2}L^3+
\frac1{L^3} \sum_{{\bf k}'} 
\sum_{m_1',m_2'} \sum_{{\bf e},{\bf e}'}
\bbra{ G({\bf k}) }_{m^{\;}_1m^{\;}_2; m_1'm_2'}
\mbra{
    e^{-{\rm i} \sbra{ {\bf k}\cdot {\bf e}-{\bf k}' \cdot {\bf e}'}} u^{m_1'm_2'}_{{\bf ee}'}
+ e^{{\rm i} \sbra{ {\bf k}\cdot{\bf e} - {\bf k} ' \cdot{\bf e}' }} u^{m_1'm_2'}_{{\bf ee}'}
}
\tilde\psi_{m_1'm_2'}({\bf k}'),
\label{eq:LS2body_boundstate}
\end{align}
\end{widetext}
where the propagator matrix $G({\bf k})$ is defined as
\begin{align}
\bbra{ G^{-1}({\bf k}) }_{m^{\;}_1m^{\;}_2;m_1'm_2'}
\equiv&~
 E\delta_{m^{\;}_1 m_1'}\delta_{m^{\;}_2 m_2'} \nonumber\\
&
 -\varepsilon_{-{\bf k} m^{\;}_1m_1'}\delta_{m^{\;}_2m_2'}
-\varepsilon_{ {\bf k} m^{\;}_2m_2'}\delta_{m^{\;}_1m_1'} \\
=& \bbra{ G^{-1} ( -{\bf k}) }_{m^{\;}_2 m^{\;}_1; m_2'm_1'} .
\end{align}
The last equality arises from the exchange symmetry of bosons.
The inverse Fourier transform, $\psi_{m_1m_2}({\bf r}) = \frac1{L^3} 
\sum_{\bf k} e^{{\rm i} {\bf k} \cdot {\bf r} } \tilde\psi_{m_1m_2}({\bf k})$, leads to
\begin{widetext}
\begin{align}
\psi_{m_1m_2}({\bf r})
= 
\phi_{m_1m_2}
+
\int_{\bf k} \sum_{m'_1, m'_2}\sum_{{\bf e}, {\bf e}'}
\bbra{
   e^{{\rm i} {\bf k} \cdot {\bf r} } \bbra{ G ({\bf k}) }_{m_1m_2;m'_1, m'_2}
+ e^{-{\rm i} {\bf k} \cdot {\bf r}} \bbra{ G ({\bf k}) }_{m_2 m_1; m'_1, m'_2}
}
e^{-{\rm i} {\bf k} \cdot {\bf e}}u_{{\bf ee}'}^{m'_1, m'_2}\psi_{m'_1, m'_2}({\bf e}'),
\label{eq:LS2body}
\end{align}
\end{widetext}
in the infinite volume limit $L \to \infty$.
Note that the properties, $[G({\bf k})]_{m_1m_2; m'_1m'_2}=[G(-{\bf k})]_{m_2m_1; m'_2m'_1}$,
 $u^{m'_1m'_2}_{{\bf ee}'}=u^{m'_2m'_1}_{{\bf ee}'}$, and $\psi_{m'_1 m'_2}({\bf r})=\psi_{m'_2 m'_1}(-{\bf r})$
are used to derive Eq.~\eqref{eq:LS2body}, which leads to a linear system of equations for the 12 unknown variables $\psi_{m_1m_2}({\bf e})$.
These 12 unknown variables and the interaction matrix elements acting on them are summarized as
\begin{align}
\psi_{+-}({\bf e}_z) &\;\Rightarrow\;
u_{{\bf e}_z{\bf e}_z}^{+-} = u_\parallel, \quad u_{(-{\bf e}_z)(+{\bf e}_z)}^{+-} = v_1, \nonumber\\
\psi_{-+}({\bf e}_z) &\;\Rightarrow\;
u_{{\bf e}_z{\bf e}_z}^{-+} = u_\parallel, \quad u_{(-{\bf e}_z)(+{\bf e}_z)}^{-+} = v_1, \nonumber\\
%\psi_{-+}({\bf e}_z) &\;\Rightarrow\;
%u_{(+{\bf e}_z)(-{\bf e}_z)}^{+-} = v_1, \nonumber\\
%\psi_{+-}({\bf e}_z) &\;\Rightarrow\;
%u_{(+{\bf e}_z)(-{\bf e}_z)}^{-+} = v_1,  \nonumber\\
\psi_{+-}({\bf e}_{x,y}) &\;\Rightarrow\;
u_{{\bf e}_{x,y}{\bf e}_{x,y}}^{+-} = u_\perp,  \nonumber\\
\psi_{-+}({\bf e}_{x,y}) &\;\Rightarrow\;
u_{{\bf e}_{x,y}{\bf e}_{x,y}}^{-+} = u_\perp,  \nonumber\\
\psi_{++}({\bf e}_{x,y}+{\bf e}_z) &\;\Rightarrow\;
u_{({\bf e}_{x,y}+{\bf e}_z)({\bf e}_{x,y}+{\bf e}_z)}^{++} = -u_\perp,  \nonumber\\
\psi_{--}({\bf e}_{x,y}-{\bf e}_z) &\;\Rightarrow\;
u_{({\bf e}_{x,y}-{\bf e}_z)({\bf e}_{x,y}-{\bf e}_z)}^{--} = -u_\perp,  \nonumber\\
%\end{align}
%\begin{align}
\psi_{++}({\bf 0}) &\;\Rightarrow\;
u_{{\bf 00}}^{++} = \mathsf{U}=\infty,  \nonumber\\
\psi_{--}({\bf 0}) &\;\Rightarrow\;
u_{{\bf 00}}^{--} = \mathsf{U}=\infty. \nonumber
\end{align}
Note that $\psi_{m_1m_2}({\bf r}) = \psi_{m_2 m _1}(-{\bf r})$ by the exchange symmetry of bosons, implying that
$\psi_{+-}(-{\bf e}_z)=\psi_{-+}({\bf e}_z)$ and
$\psi_{-+}(-{\bf e}_z)=\psi_{+-}({\bf e}_z)$.
For concreteness, Eq.~\eqref{eq:LS2body} can be expressed as
\begin{align}
\mathcal{I} 
{\bm \psi} =&~
{\bm \phi}+
 \mathcal{A}(E) {\bm \psi}
\quad \Rightarrow \quad {\bm \psi} = \bbra{ \mathcal{I} - \mathcal{A}(E) }^{-1} {\bm \phi}
\label{eq:numeq}
 ,
 \end{align}
with
%\begin{widetext}
 \begin{align}
 {\bm \psi}^{\rm t} =&\big[
 \psi_{+-}({\bf e}_z) ,
\psi_{-+}({\bf e}_z) ,
\psi_{+-}({\bf e}_{x}) ,
\psi_{+-}({\bf e}_{y}) ,\nonumber\\
&~\psi_{-+}({\bf e}_{x}) ,
\psi_{-+}({\bf e}_{y}) , 
\psi_{++}({\bf e}_{x}+{\bf e}_z),
\psi_{++}({\bf e}_{y}+{\bf e}_z),\nonumber\\
&~\psi_{--}({\bf e}_{x}-{\bf e}_z) ,
\psi_{--}({\bf e}_{y}-{\bf e}_z) , 
\mathsf{U} \psi_{++}({\bf 0}) , 
\mathsf{U} \psi_{--}({\bf 0})
 \big],\nonumber\\
{\bm \phi}^{\rm t} =&\big[
\phi_{+-},
\phi_{-+},
\phi_{+-},
\phi_{+-},
\phi_{-+},
\phi_{-+},\nonumber\\
&~\phi_{++},
\phi_{++},
\phi_{--},
\phi_{--},
\phi_{++}, 
\phi_{--}
 \big],\nonumber
\end{align}
%\end{widetext}
where $\mathcal{I}$ and $\mathcal{A}(E)$ are 12 by 12 matrices;
$\mathcal{I}_{\nu \nu'}= \delta_{\nu \nu'} (1- \delta_{\nu 11}- \delta_{\nu 12})$, and components of $\mathcal{A}(E)$ are integrals over ${\bf k}$-space.
The integration in Eq.~\eqref{eq:LS2body} is performed by applying the Gaussian quadrature rule to discretize ${\bf k}$-integrals.
It is worth noting that $u^{\pm\pm}_{\bf 00} \psi_{\pm\pm}({\bf 0})$ is finite for the self-consistent solution, while $\psi_{\pm\pm}({\bf 0})=0$ because of $u^{\pm\pm}_{\bf 00} = \infty$.

The substitution of the obtained ${\bm \psi}$ to Eq.~\eqref{eq:LS2body} provides the wave function for the two-magnon scattering.
The value of $a$ is extracted from the wave function by multiplying both sides of Eq.~\eqref{eq:LS2body} by $\phi^*_{m_1m_2}$ and taking the sum over $m_1$ and $m_2$:
\begin{widetext}
\begin{align}
&\sum_{m_1,m_2}\phi^*_{m_1m_2}\psi_{m_1m_2}({\bf r}) \nonumber\\
&= 1+
\int_{\bf k} \sum_{m_1, m_2,m'_1, m'_2}\sum_{{\bf e}, {\bf e}'}
\phi^*_{m_1m_2}
\bbra{
   e^{{\rm i} {\bf k} \cdot {\bf r} } \bbra{ G ({\bf k}) }_{m_1m_2;m'_1, m'_2}
+ e^{-{\rm i} {\bf k} \cdot {\bf r}} \bbra{ G ({\bf k}) }_{m_2 m_1; m'_1, m'_2}
}
e^{-{\rm i} {\bf k} \cdot {\bf e}}u_{{\bf ee}'}^{m'_1, m'_2}\psi_{m'_1, m'_2}({\bf e}'),\nonumber\\
&\xrightarrow[]{\overline{r}  \to \infty}
1-\frac{a}{\overline{r}},
\label{eq:two-magnon-wave-function}
\end{align}
\end{widetext}
where 
$\overline{r} = |\overline{\bf r}|$ and
$\overline{\bf r} = \sbra{ \sqrt{\frac{m_x}{m_z}} r_x, \sqrt{\frac{m_x}{m_z}}r_y, \frac{r_z}{2} }$
with ${\bf r} = r_x {\bf e}_x + r_y {\bf e}_y + r_z {\bf e}_z$.
The divergent behavior of the Green's function in the infrared limit,
\begin{align}
\sum_{m_1, m_2} \phi^*_{m_1m_2} \bbra{ G ({\bf k}) }_{m_1m_2;m'_1, m'_2}
\xrightarrow[]{\overline{\bf k}  \to 0}
-\frac{m_z}{ \overline{\bf k}^2 } \delta_{m_1 m_1'} \delta_{m_2 m_2'},
\end{align}
determines the asymptotic behavior of the integral in Eq.~\eqref{eq:two-magnon-wave-function}.
The definition of $\overline{{\bf k}}$ is given in Eq.~\eqref{eq:Disprel} of the main text:
\begin{align}
\overline{{\bf k}} =&~( \overline{k}_x, \overline{k}_y, \overline{k}_z ) \nonumber\\
=& \bbra{
\sqrt{\frac{m_z}{m_x}} \sbra{k_1 -\frac{k_3}{2}},
\sqrt{\frac{m_z}{m_x}} \sbra{k_2 -\frac{k_3}{2}},
k_3}.
\end{align}
After changing the variables and extending the integration interval to $[-\infty , \infty]$, we obtain
the following asymptotic behavior for the integral in Eq.~\eqref{eq:two-magnon-wave-function}:
\begin{align}
\int_{\bf k} \frac{ e^{{\rm i} {\bf k} \cdot {\bf r}} }{ \overline{\bf k}^2 } 
=& \frac{1}{2\pi} \frac{m_x}{m_z}
 \int_{ - \sqrt{ \frac{m_z}{m_x}} \pi}^{\sqrt{ \frac{m_z }{m_x}} \pi} d\overline{k}_x
  \int_{ - \sqrt{ \frac{m_z}{m_x}} \pi}^{\sqrt{ \frac{m_z}{ m_x}} \pi} d\overline{k}_y
   \int_{ - \pi}^{\pi} d\overline{k}_z
   \frac{ e^{{\rm i} \overline{\bf k} \cdot \overline{\bf r}} }{ \overline{\bf k}^2 } \nonumber\\
\simeq& \frac{m_x}{m_z} \frac{1}{4\pi \overline{r}},
\end{align}
which leads to a simple expression for the $s$-wave scattering length,
\begin{align}
a =&~
\frac{m_x}{4\pi}
\sum_{m_1, m_2}
\sum_{{\bf e}, {\bf e}'}
\phi^*_{m_1m_2}
\mbra{
\psi_{m_1m_2} ({\bf e}') 
+
\psi_{m_2m_1} ({\bf e}') 
}
u^{m_1 m_2}_{{\bf ee}'}
\nonumber\\
=&~
\frac{m_x}{4\pi}
\sum_{m_1, m_2}
\sum_{{\bf e}, {\bf e}'}
\psi_{m_1m_2} ({\bf e}') 
u^{m_1 m_2}_{{\bf ee}'}
.
\end{align}
The second line is obtained from the first one by using $\phi_{m_1m_2} = 1/2$.

Finally, for the two-magnon bound states, we set $E < 2 \mathcal{E}_0$ and $\phi_{m_1m_2}=0$ in Eq.~\eqref{eq:LS2body}.
Then, we can obtain the two-magnon bound state energy $E$ by numerically solving $\det \bbra{\mathcal{I}-\mathcal{A}(E) } = 0$ instead of Eq.~\eqref{eq:numeq}.
Its binding energy is provided by $\Delta = E - 2 \mathcal{E}_0$.

\subsection{Three-magnon problem}\label{sec:three_magnon_problem}
The Lippmann-Schwinger equation for the three-magnon problem is derived in the same way as in the previous cases.
We introduce the real space representation of the  three-magnon wave function and its Fourier transform,
\begin{align}
\Psi({\bf r}_1,{\bf r}_2,{\bf r}_3)=&
\bracketii{\emptyset}{a_{{\bf r}_3} a_{{\bf r}_2} a_{{\bf r}_1} }{\Psi},%\\
\end{align}
\begin{align}
\tilde{\Psi}_{ {\bf m}
%m_1m_2m_3
}({\bf k}_1,{\bf k}_2,{\bf k}_3)=&
\Big[
\prod_{\nu = 1}^3 \sum_{{\bf r}_\nu \in m_\nu } 
\Big]
 e^{- {\rm i} \sum_{\nu' =1}^3{\bf k}_{\nu'} \cdot {\bf r}_{\nu'} }
\Psi({\bf r}_1,{\bf r}_2,{\bf r}_3),\\
\psi_{
{\bf m}
%m_1m_2;m_3
}({\bf r};{\bf k} ) =&
\frac{1}{L^3} 
\sum_{{\bf k}'}
e^{{\rm i} {\bf k}' \cdot  {\bf r} } \tilde{\Psi}_{
%m_1m_2m_3
{\bf m}
} (-{\bf k} -{\bf k}', {\bf k}', {\bf k}) \nonumber\\
=&
e^{- {\rm i} {\bf k} \cdot {\bf r} } \psi_{
(m_2 , m_1 , m_3)
%m_2 m_1;m_3
} (-{\bf r};{\bf k}),
\end{align}
with ${\bf m} = (m_1,m_2,m_3)$.
Note that the zero center-of-mass momentum condition, ${\bf k}_1+{\bf k}_2+{\bf k}_3 = {\bf 0}$, and the exchange symmetry of bosons are imposed in the above transformation.
In this way, we obtain a linear set of equations for the three-magnon problem:
\begin{widetext}
\begin{align}
& \psi_{
{\bf m}
%m^{\;}_1m^{\;}_2;m^{\;}_3
}({\bf r};{\bf k}) 
= \int_{{\bf k}'}\sum_{m'_1,m'_2,m'_3}\sum_{{\bf e},{\bf e}'}
\Big[ \nonumber\\
&\quad\quad \;\left.
\mbra{ 
e^{{\rm i}{\bf k}' \cdot{\bf r}} \bbra{ G({\bf k}' ,{\bf k} )}_{m^{\;}_1m^{\;}_2m^{\;}_3;m'_1m'_2m'_3}
+ e^{-{\rm i}({\bf k}' +{\bf k} )\cdot {\bf r}} \bbra{ G ({\bf k}' ,{\bf k} )}_{m^{\;}_2m^{\;}_1m^{\;}_3;m'_1m'_2m'_3}
}
e^{-{\rm i}{\bf k}' \cdot {\bf e}}u_{{\bf ee}'}^{m'_1m'_2}\psi_{
%m'_1m'_2;m'_3
(m'_1,m'_2,m'_3)
}({\bf e}';{\bf k} ) \right. \nonumber\\
&\quad
+
\left[
\mbra{
e^{{\rm i}{\bf k}' \cdot{\bf r}} \bbra{ G({\bf k}' ,{\bf k} ) }_{m^{\;}_1m^{\;}_2m^{\;}_3;m'_1m'_2m'_3}
+ e^{-{\rm i}({\bf k}' +{\bf k} )\cdot{\bf r}} \bbra{G({\bf k}' ,{\bf k} ) }_{m^{\;}_2m^{\;}_1m^{\;}_3;m'_1m'_2m'_3}
}
e^{-{\rm i}{\bf k} \cdot {\bf e}} 
\right. \nonumber\\
&\quad + 
\left. \left.
\;\mbra{
e^{{\rm i}{\bf k}' \cdot{\bf r}} \bbra{G({\bf k}' ,{\bf k} )} _{m^{\;}_1m^{\;}_2m^{\;}_3;m'_3m'_2m'_1}
+ e^{-{\rm i}({\bf k}' +{\bf k} )\cdot{\bf r}} \bbra{G({\bf k}' ,{\bf k} )} _{m^{\;}_2m^{\;}_1m^{\;}_3;m'_3m'_2m'_1}
}
e^{{\rm i}({\bf k}' +{\bf k} )\cdot{\bf e}}
\right]
u_{{\bf e}{\bf e}'}^{m'_1m'_3}\psi_{
%m'_1m'_3;m'_2
(m'_1,m'_3,m'_2)
}({\bf e}';{\bf k}' )
\right],
\label{eq:LS3body}
\end{align}
\end{widetext}
where the propagator matrix $G({\bf k}',{\bf k})$ is defined as
\begin{align}
& \bbra{G^{-1}({\bf k}' ,{\bf k} ) }_{m^{\;}_1m^{\;}_2m^{\;}_3;m'_1m'_2m'_3} \nonumber\\
&\equiv E\delta_{m^{\;}_1m'_1}\delta_{m^{\;}_2m'_2}\delta_{m^{\;}_3m'_3}
- \varepsilon_{-({\bf k}' +{\bf k} )m^{\;}_1m'_1}  \delta_{m^{\;}_2m'_2}\delta_{m^{\;}_3m'_3}\nonumber\\
&\quad
- \varepsilon_{{\bf k}' m^{\;}_2m'_2}                  \delta_{m^{\;}_3m'_3}\delta_{m^{\;}_1m'_1}
- \varepsilon_{{\bf k} m^{\;}_3m'_3}                   \delta_{m^{\;}_1m'_1}\delta_{m^{\;}_2m'_2}.
\end{align}
The 24 unknown functions and the interaction matrix elements acting on them are summarized as
\begin{align}
\psi_{
%+-;m
(+,-,m)
}({\bf e}_z;{\bf k}) &\;\Rightarrow\;
u_{{\bf e}_z{\bf e}_z}^{+-} = u_\parallel, \; u_{(-{\bf e}_z)(+{\bf e}_z)}^{+-} = v_1, \nonumber\\
\psi_{
%-+;m
(-,+,m)
}({\bf e}_z;{\bf k}) &\;\Rightarrow\;
u_{{\bf e}_z{\bf e}_z}^{-+} = u_\parallel, \; u_{(-{\bf e}_z)(+{\bf e}_z)}^{-+} = v_1, \nonumber\\
%\psi_{+-;m}(-{\bf e}_z;\k) %=e^{i\k\cdot{\bf e}_z}
%\psi_{-+;m}({\bf e}_z;{\bf k}) 
%&\;\Rightarrow\;
%u_{(+{\bf e}_z)(-{\bf e}_z)}^{+-} = v_1,  \nonumber\\
%\psi_{-+;m}(-{\bf e}_z;\k)%=e^{i\k\cdot{\bf e}_z}
%\psi_{+-;m}({\bf e}_z;{\bf k}) 
%&\;\Rightarrow\;
%u_{(+{\bf e}_z)(-{\bf e}_z)}^{-+} = v_1,  \nonumber\\
\psi_{
%+-;m
(+,-,m)
}({\bf e}_{x,y};{\bf k}) &\;\Rightarrow\;
 u_{{\bf e}_{x,y}{\bf e}_{x,y}}^{+-} = u_\perp,  \nonumber\\
\psi_{
%-+;m
(-,+,m)
}({\bf e}_{x,y};{\bf k}) &\;\Rightarrow\;
u_{{\bf e}_{x,y}{\bf e}_{x,y}}^{-+} = u_\perp,  \nonumber\\
\psi_{
%++;m
(+,+,m)
}({\bf e}_{x,y}+{\bf e}_z;{\bf k}) &\;\Rightarrow\;
u_{({\bf e}_{x,y}+{\bf e}_z)({\bf e}_{x,y}+{\bf e}_z)}^{++} = -u_\perp,  \nonumber\\
\psi_{
%--;m
(-,-,m)
}({\bf e}_{x,y}-{\bf e}_z;{\bf k}) &\;\Rightarrow\;
u_{({\bf e}_{x,y}-{\bf e}_z)({\bf e}_{x,y}-{\bf e}_z)}^{--} = -u_\perp,  \nonumber\\
\psi_{
%++;m
(+,+,m)
}({\bf 0};{\bf k}) &\;\Rightarrow\;
u_{{\bf 00}}^{++} = \mathsf{U} = \infty,  \nonumber\\
\psi_{
%--;m
(-,-,m)
}({\bf 0};{\bf k}) &\;\Rightarrow\;
u_{\bf 00}^{--} = \mathsf{U} = \infty, \nonumber
\end{align}
where $m=\pm$.
Note that the exchange symmetry of bosons implies
$\psi_{
%m_1m_2;m_3
(m_1,m_2,m_3)
}({\bf r};{\bf k})=e^{-{\rm i} {\bf k} \cdot {\bf r} } 
\psi_{
%m_2m_1;m_3
(m_2,m_1,m_3)
}(-{\bf r};{\bf k})$ 
and thus,
$\psi_{
%+-;m
(+,-,m)
}(-{\bf e}_z;{\bf k})=e^{{\rm i}{\bf k}\cdot{\bf e}_z}
\psi_{
%-+;m
(-,+,m)
}({\bf e}_z;{\bf k})$, and
$\psi_{
%-+;m
(-,+,m)
}(-{\bf e}_z;{\bf k})=e^{{\rm i}{\bf k}\cdot{\bf e}_z}
\psi_{
%+-;m
(+,-,m)
}({\bf e}_z;{\bf k})$.
 Similar to the two-magnon problem, $u^{\pm\pm}_{\bf 00} \psi_{
%\pm\pm;m
(\pm,\pm,m)
}({\bf 0};{\bf k})$ is assumed to be finite, while 
$\psi_{
%\pm\pm;m
(\pm,\pm,m)
}({\bf 0};{\bf k})=0$. 
The self-consistency of this assumption is confirmed by the numerical solutions of Eq.~\eqref{eq:LS3body}.

To reduce the computational cost, we exploit the symmetry of the wave function $\psi_{\bf m}({\bf r};{\bf k})$.
The propagator has the following symmetry properties inherited from the effective boson Hamiltonian $\mathcal{H}_{\rm eff}$:
%\begin{widetext}
\begin{align}
&\bbra{G({\bf k}',{\bf k})}_{m^{\;}_1m^{\;}_2 m^{\;}_3; m'_1 m'_2 m'_3} \nonumber\\
&=
 \bbra{G({\bf k}',{\bf k})}_{m^{\;}_1m^{\;}_2 m^{\;}_3; m'_1 m'_2m'_3} \big|_{ \tilde{k}_x\leftrightarrow \tilde{k}_y, \tilde{k}'_x\leftrightarrow \tilde{k}'_y} \nonumber\\
&=
 \bbra{G({\bf k}',{\bf k})}_{\overline{m^{\;}_1} \; \overline{m^{\;}_2} \; \overline{m^{\;}_3}; \overline{m'_1} \; \overline{m'_2} \; \overline{m'_3}} \big|_{
\begin{subarray}{l}
\tilde{k}_x\to-\tilde{k}_x,\tilde{k}_y\to-\tilde{k}_y,\\
\tilde{k}'_x\to-\tilde{k}'_x,\tilde{k}'_y\to-\tilde{k}'_y
\end{subarray}
} \nonumber\\
&=
 \bbra{G({\bf k}',{\bf k})}_{\overline{m^{\;}_1} \; \overline{m^{\;}_2} \; \overline{m^{\;}_3}; \overline{m'_1} \; \overline{m'_2} \; \overline{m'_3}} \big|_{{\tilde{k}}_z\to-{\tilde{k}}_z,{\tilde{k}}'_z\to-{\tilde{k}}'_z} {,}
\end{align}
%\end{widetext}
with ${\bf k}^{(\prime)}= \frac{1}{2\pi} \bbra{  \tilde{k}_x^{(\prime)} \tilde{\bf G}_x  + \tilde{k}_y^{(\prime)} \tilde{\bf G}_y  + \frac{\tilde{k}_z^{(\prime)}}2 \tilde{\bf G}_z  }$.
By applying these symmetries on the Lippmann-Schwinger equation for the three-magnon problem, we can demonstrate that
\begin{align}
& \psi_{
{\bf m}
%m^{\;}_1m^{\;}_2;m^{\;}_3
}({\bf r};{\bf k}) ,\quad
\psi_{
{\bf m}
%m^{\;}_1m^{\;}_2;m^{\;}_3
}({\bf r};{\bf k})|_{r_x\leftrightarrow r_y,{\tilde{k}}_x\leftrightarrow {\tilde{k}}_y}, \nonumber\\
&\psi_{
{\overline{\bf m}}
%\overline{ m^{\;}_1} \;  \overline{ m^{\;}_2}; \overline{ m^{\;}_3}
 }({\bf r};{\bf k})|_{r_x\to-r_x,r_y\to-r_y,{\tilde{k}}_x\to-{\tilde{k}}_x,{\tilde{k}}_y\to-{\tilde{k}}_y} , \nonumber\\
&\psi_{
{\overline{\bf m}}
%\overline{ m^{\;}_1} \;  \overline{ m^{\;}_2}; \overline{ m^{\;}_3} 
}({\bf r};{\bf k})|_{r_z\to-r_z,{\tilde{k}}_z\to-{\tilde{k}}_z}
\end{align}
with $\overline{\bf m}=
(\overline{ m^{\;}_1} ,  \overline{ m^{\;}_2}, \overline{ m^{\;}_3})$
satisfy the same integral equation. For a non-degenerate eigenstate, the wavefunction must take the symmetric form,
\begin{align}
 \psi_{
 {\bf m}
 %m^{\;}_1m^{\;}_2;m^{\;}_3
 }({\bf r};{\bf k}) =&~ \pm
\psi_{
{\bf m}
%m^{\;}_1m^{\;}_2;m^{\;}_3
}({\bf r};{\bf k})|_{r_x\leftrightarrow r_y,{\tilde{k}}_x\leftrightarrow {\tilde{k}}_y} \nonumber\\
=&~ \pm
\psi_{
{\overline{\bf m}}
%\overline{ m^{\;}_1} \;  \overline{ m^{\;}_2}; \overline{ m^{\;}_3} 
}({\bf r};{\bf k})|_{r_x\to-r_x,r_y\to-r_y,{\tilde{k}}_x\to-{\tilde{k}}_x,{\tilde{k}}_y\to-{\tilde{k}}_y} \nonumber\\
=&~ \pm
\psi_{
{\overline{\bf m}}
%\overline{ m^{\;}_1} \;  \overline{ m^{\;}_2}; \overline{ m^{\;}_3} 
}({\bf r};{\bf k})|_{r_z\to-r_z,{\tilde{k}}_z\to-{\tilde{k}}_z}.
\label{eq:symmetry3body}
\end{align}
In the calculation, we choose the ``+'' sign  for each symmetry operation because the Efimov states belong to the $s$-wave sector.
As in the case of the  two-magnon problem, we apply the Gaussian quadrature rule to discretize ${\bf k}$-integrals.
A bound state with a larger characteristic size (closer to the three-magnon continuum) requires finer momentum space discretization. For the finest momentum space discretization that can be achieved with current supercomputers, we can obtain the energy eigenvalues and the corresponding wave functions for the two lowest energy states in the Efimov tower.

\section{Wave function of the Efimov state in the unitary limit}\label{sec:wave_function_of_Efimov_state}

\begin{figure}[t]
\includegraphics[width= 0.95 \columnwidth]{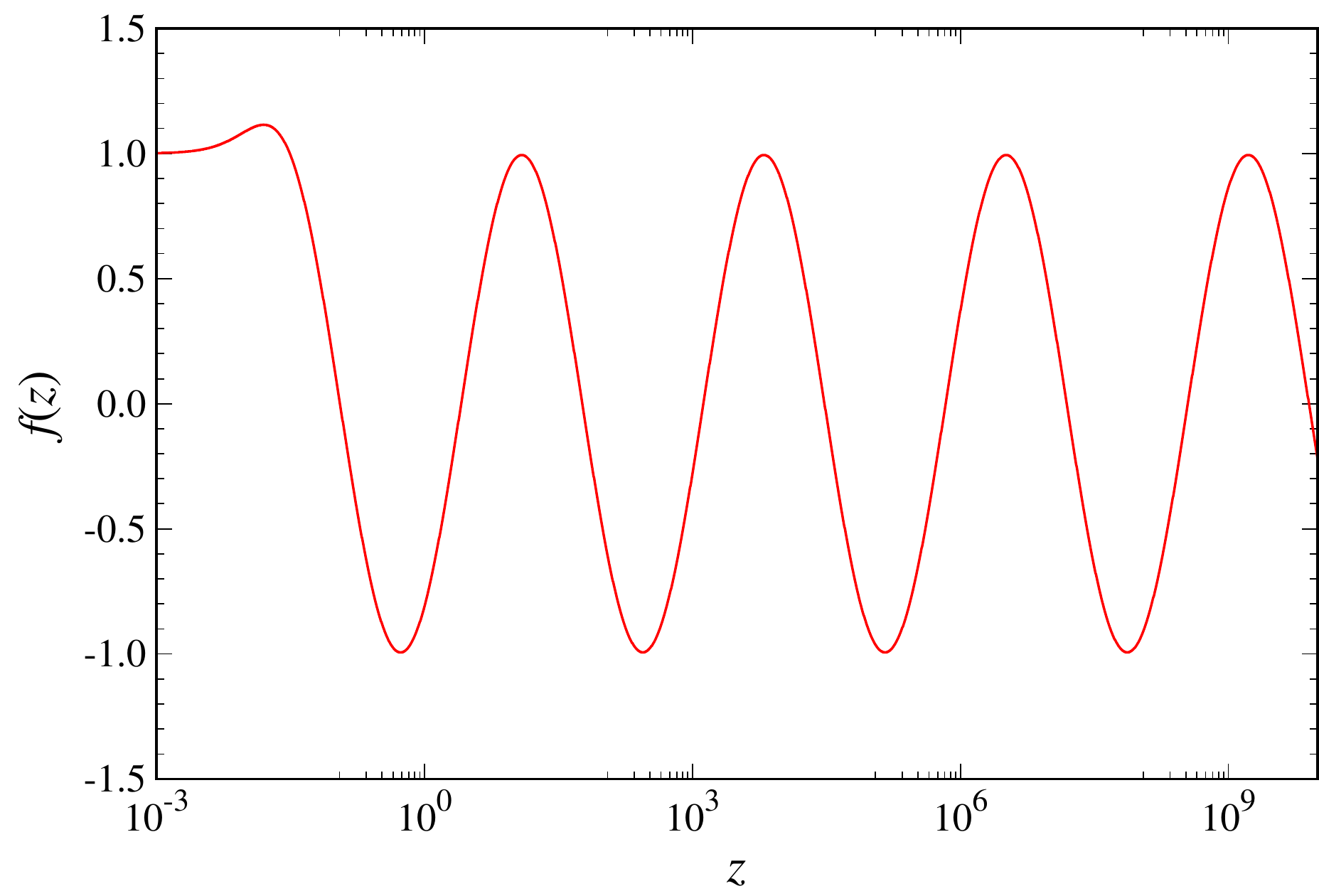}
\caption{
Universal function $f(z)$ associated with  Efimov states.
The normalization is chosen to satisfy $f(0)=1$.
\label{fig:fz}
}
\end{figure}

In this section, we consider the wave function of the Efimov state in continuum space~\cite{Braaten:2006}.
The three-body bound state problem for bosons with an isotropic mass $m$ can be reduced to a solution of the following Skorniakov-Ter-Martirosian (STM) integral equation:
\begin{align}
&\left[
\sqrt{\frac{3{\bf p}^2}{4}-mE-{\rm i}0^+} - \frac1a
\right]\tilde\phi({\bf p}) \nonumber\\
%= \int\!\frac{d{\bf q}}{(2\pi)^3}\frac{8\pi}{{\bf p}^2+{\bf q}^2+ {\bf p}\cdot{\bf q}-mE-{\rm i}0^+}\tilde\phi({\bf q}),
&= \frac{1}{\pi^2} \int\!d{\bf q} \frac{1}{{\bf p}^2+{\bf q}^2+ {\bf p}\cdot{\bf q}-mE-{\rm i}0^+}\tilde\phi({\bf q}),
\end{align}
where ${\bf p}$ is the momentum corresponding to $\overline{\bf k}$ in Eq.~\eqref{eq:Disprel} and the wave function $\tilde\phi({\bf p}) \propto \int d{\bf R} \int d{\bf r}\,e^{-{\rm i} {\bf p}\cdot {\bf r} }\phi({\bf R}-{\bf r}/2,{\bf R}-{\bf r}/2,{\bf R}+{\bf r}/2)$
corresponds to $\psi_{{\bf m}}({\bf e}; {\bf k})$ in Eq.~\eqref{eq:essential_portion} of the main text.
By setting $E\equiv-\frac{\kappa^2}{m}$ and $\tilde\phi({\bf p})=\tilde\phi(p=|{\bf p}|)$, we obtain
\begin{align}
&\bbra{
\sqrt{\frac{3p^2}{4}+\kappa^2} - \frac1a
}
\tilde\phi(p) \nonumber\\
&= \frac2\pi\int_0^\infty\!dq\,\frac{q}{p}
\ln\!\left(\frac{p^2+q^2+pq+\kappa^2}{p^2+q^2-pq+\kappa^2}\right)\tilde\phi(q).
\end{align}
In the unitary limit $a\to\infty$, the $n$-th bound state solution is given by
\begin{align}
\lim_{n\to\infty}\kappa_n = \frac{\kappa_*}{\lambda^n}
\qquad\text{with}\qquad
\lim_{n\to\infty}\tilde\phi_n(p) = \frac{f\left(\sqrt{\frac34}\frac{p}{\kappa_n}\right)}{\frac34\left(\frac{p}{\kappa_n}\right)^2+1},
\end{align}
where
\begin{align}
f(z) = \frac{\sin[s_0\mathrm{arcsinh}(z)]}{s_0z}\sqrt{z^2+1}
\end{align}
is the universal function plotted in Fig.~\ref{fig:fz}~\cite{Gogolin:2008}.
The constant $s_0 = 1.00624$ solves
\begin{align}
\frac{8}{\sqrt{3} s_0} \frac{ \sinh \sbra{ \frac{\pi}6 s_0 } }{ \cosh \sbra{ \frac{\pi}2 s_0 } } = 1.
\end{align}

\begin{figure}[t]
  \centering
 \includegraphics[trim=22 0 20 0,clip,width=\columnwidth]{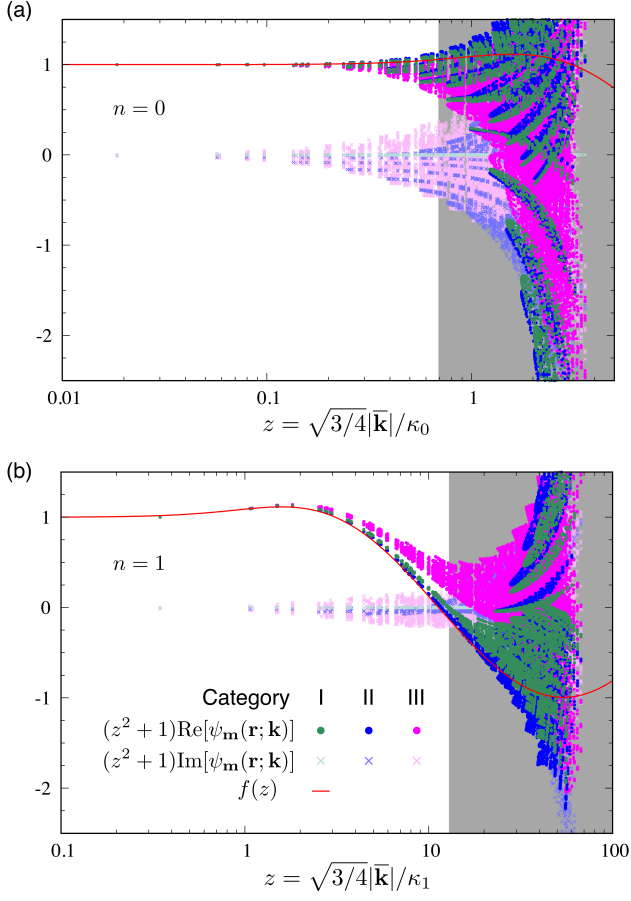}
  \caption{
Wave functions of the two lowest three-magnon bound states at the critical field $\mu_0H_c = 12.91$~T.
The universal function $f(z)$ and the numerical solutions for (a) the ground state ($n=0$) and for (b) the first excited state ($n=1$) are compared as functions of the rescaled wave number $z = \sqrt{3/4}|\overline{\bf k}|/\kappa_n$ normalized by $\kappa_n$ for each $n$.
All the calculated 24 functions are grouped into three categories, I, II, and III, and 
the common symbols are used for each group as shown in the legend.
The gray shaded regions indicate the nonuniversal regime ($|\overline{\bf k}|>1$).
} 
  \label{fig:result2b}
\end{figure}

In Fig.~\ref{fig:result2} of the main text, we demonstrate that the universal function $f(z)$ well describes
the three-magnon wave functions especially for the $n=1$ state.
Figure~\ref{fig:result2b} now shows 
the full comparison between the universal function and solutions of the Lippmann-Schwinger equation [Eq.~\eqref{eq:LS3body}]. 
Our numerical calculation provides a set of 24 functions as a solution, which are grouped into three categories as
\begin{align*}
{\rm I}:&~\psi_{(m,m,m')} ({\bf 0}; {\bf k}),\\ % x 4
{\rm II}:&~\psi_{(m,\overline{m},m')} ({\bf e}_z; {\bf k}), \\% x 4
{\rm III}:&~\psi_{(m,\overline{m},m')} ({\bf e}_{x,y}; {\bf k}),% x 8
~\psi_{(m,m,m')} ({\bf e}_{x,y}+ m{\bf e}_z; {\bf k}).% x 8
\end{align*} 
In addition to the category II presented in Fig.~\ref{fig:result2}, the wave functions of categories I and III are also plotted in Fig.~\ref{fig:result2b}.
All the wave functions show good agreement with the universal function especially for the $n=1$ state.

%%%%%%%%%%%%%%%%%%%%%%%

%\bibliographystyle{apsrev}

\bibliography{draft} 

\end{document}